\documentclass[11pt,a4paper]{article}
\usepackage{jstyle}
\usepackage{hyperref,slashed,empheq}
\usepackage{framed}
\usetikzlibrary{calc,decorations.markings}

\newcommand{\Y}{\mathbb{Y}}
\newcommand{\N}{\mathbb{N}}

\newcommand{\dd}{\mathrm{d}}

\newcommand{\Pd}[1]{\mathcal{P}_{#1}}

\definecolor{mygreen}{rgb}{0.32,1.65,0.50}
\definecolor{myblue}{rgb}{0.08, 0.38, 0.74}
\definecolor{theRed}{rgb}{0.56,0,0}

\author[a]{Thomas BASILE}
\author[a]{\quad Euihun JOUNG}
\author[b]{\quad Shailesh LAL}
\author[a]{\quad Wenliang LI}

\affiliation[a]{Department of Physics and Research Institute of Basic
  Science, \\ Kyung Hee University,\\ Seoul 02447, Korea}
\affiliation[b]{Centro de Fisica do Porto e Departamento de Fisica e
  Astronomia \\ Faculdade de Ciencias da Universidade do
  Porto,\\ Rua do Campo Alegre 687, 4169-007 Porto, Portugal}

\emailAdd{thomas.basile@khu.ac.kr}
\emailAdd{euihun.joung@khu.ac.kr}
\emailAdd{slal@fc.up.pt}
\emailAdd{lii.wenliang@gmail.com}

\abstract{The zeta function of an arbitrary field in
  $(d+1)$-dimensional anti-de Sitter (AdS) spacetime 
  is expressed
  as an integral transform of the corresponding $so(2,d)$
  representation character, thereby extending the results of
  \href{https://arxiv.org/abs/1603.05387}{\tt [1603.05387]} for
  AdS$_4$ and AdS$_5$ to arbitrary dimensions. 
  The integration in the variables associated with the $so(d)$ part
  of the character
  can be recast into a more explicit form using derivatives.
  The explicit derivative expressions are presented for 
  AdS$_{d+1}$ with $d=2,3,4,5,6$. }

\begin{document}

\title{\centering {\huge Character Integral Representation\\ of Zeta
    function in AdS$_{d+1}$}:\\
  \bigskip
  \Large I. Derivation of the general formula}

\maketitle

\section{Introduction}
\label{sec:intro}

The one-loop free energy is one of the simplest physical quantities
capturing non-trivial quantum effects.  It is divergent due to the
contribution from the modes having infinite energy but can be neatly
regularized by making use of the spectral zeta function, namely in the
scheme of the zeta function regularization.  The zeta function for 
any field (massive or massless, and of arbitrary spin)
 in four-dimensional Anti-de-Sitter (AdS)
spacetime has been first calculated by Camporesi and Higuchi in
\cite{Camporesi:1993mz} and generalized to higher dimensions by the
same authors in \cite{Camporesi:1994ga, Camporesi1994}.

Zeta functions in AdS are useful in the test of certain AdS/CFT
dualities: the one-loop free energy or vacuum energy in AdS spacetime
corresponds to the non-planar contribution of the CFT free energy on
the boundary. Since the typical AdS theories under consideration
contain infinitely many fields, computing their one-loop free energy
is a non-trivial task.\footnote{If instead one considers the change in
  free energy by taking an alternative boundary condition for one of
  the AdS fields, this technical difficulty does not arise.  This
  quantity is matched to the double-trace deformation of the
  corresponding CFT \cite{Witten:2001ua, Berkooz:2002ug,
    Gubser:2002zh, Gubser:2002vv, Diaz:2007an}.}  When the AdS theory
is a higher-spin gravity, the one-loop free energy is calculable in an
analytic manner, even though the field content still contains
infinitely many elements.  An interesting observation made in
\cite{Giombi:2013fka,Giombi:2014iua} is that the summation of the zeta
functions over the field content is convergent while that of the
regularized one-loop free energy is divergent. This is interesting as
it signifies that the zeta function regularization renders finite both
the high energy divergence and the spectrum sum divergence.

The viability of the one-loop free energy computation in higher-spin
gravities heavily relies on the simple structure of the spectrum: e.g.
in the case of the non-minimal type-A theory, first
  constructed in four dimensions \cite{Vasiliev:1990en,
    Vasiliev:1992av} and later extended to arbitrary dimensions in
  \cite{Vasiliev:2003ev}, the spectrum consists of massless fields
of all integer spins.  The summation over a field content becomes
quite cumbersome if the content itself does not have a simple
expression. This kind of difficulty was encountered in the computation
of the one-loop free energy of the AdS fields dual to the operators
tri- and quadri-linear in free conformal scalar fields
\cite{Bae:2016rgm}.  In the absence of the single-trace condition ---
that is, the cyclic projection on the operators --- the field content
could be expressed in a few lines, which could be used to calculate
the one-loop free energy although it required quite burdensome works.
What is worse is that the field content with cyclic projection does
not have any manageable expression, hence it seems impossible to
proceed in this way.

A key observation to bypass this problem is that the field content of
an AdS theory dual to a free CFT can be derived group theoretically.
One of the most efficient and general methods for such a derivation is
the use of Lie algebra character.  In fact, the spectrum of the
four-dimensional type-A higher-spin gravity was obtained using the
$so(2,3)$ character --- namely, the Flato-Fronsdal theorem
\cite{Flato:1978qz} (later generalized in arbitrary dimensions in
\cite{Angelopoulos:1999bz, Vasiliev:2004cm}).  Since both the zeta
function and the character are determined uniquely by the labels of
$so(2,d)$ representations, we can devise a linear map which send the
character of an $so(2,d)$ representation to the zeta function of the
AdS field carrying the same representation.  If the linear map itself
does not depend on the labels of the representation, we can use it for
the character of a reducible representation without decomposing it
into irreducible pieces.  Such a map was explicitly constructed for
bosonic fields in AdS$_4$ and any fields in AdS$_5$ and was named as
``character integral representation of zeta function (CIRZ)'' in
\cite{Bae:2016rgm}.  The CIRZ turned out to be very efficient in
evaluating the one-loop free energies.  Notably, the computation of
the non-minimal type-A higher-spin gravity becomes almost trivial.
Moreover, the CIRZ allowed to tackle the
  one-loop free energy computation of the stringy AdS theory dual to
  free matrix model CFTs in the $N\to \infty$ limit 
  \cite{Bae:2016rgm,Bae:2016hfy,Bae:2017spv,Bae:2017fcs}. 
   The CIRZ also proved useful in other vector
model dualities: the ``colored'' higher-spin gravity (where the
four-dimensional CIRZ was generalized to fermionic fields)
\cite{Pang:2016ofv} and the type-J higher-spin theories (whose
  conjectured dual are free vector model based on a massless spin-$j$
  field) \cite{Bae:2016xmv}.
Some key elements of the CIRZ were also used in \cite{Skvortsov:2017ldz}.

In this paper, we aim to derive the CIRZ in dimensions different from
AdS$_4$ and AdS$_5$. More precisely, we seek for the formula extending
the CIRZ in AdS$_{d+1}$ with arbitrary integer $d$ $\geq 2$.  Although
the dimensional dependence in higher-spin gravity is rather minimal,
most of the results in the literature \cite{Giombi:2013fka,
  Giombi:2014iua, Giombi:2014yra, Giombi:2016pvg, Gunaydin:2016amv,
  Brust:2016xif} concern only specific dimensions for technical
reasons, except in \cite{Skvortsov:2017ldz} where the results of the
type-A higher-spin gravity are extended to arbitrary (non-integer)
dimensions. From the viewpoint of physical applications, one might not
need to care about higher dimensions yet, but it is at the same time
tempting to obtain results with parametric dependence on $d$.  As
usual, generalities may provide new and valuable lessons on what is
considered to be well-understood.  It is actually the case here for
the CIRZ: in the course of its derivation for general dimensions, we
find many new insights on the zeta function, $so(2,d)$ character and
the relations between them.

The general CIRZ formula we obtain is an integral transformation of
$so(2,d)$ character and has a quite simple structure: it is in a sense
even simpler than the original CIRZ expressions obtained in AdS$_4$
and AdS$_5$ in \cite{Bae:2016rgm}. In order to deliver a flavor of our
results, let us write down the expressions of the general CIRZ derived
in \hyperref[sec:CIRZ]{Section \ref{sec:CIRZ}}. Firstly, for odd AdS
dimensions $d+1=2r+1$ (or equivalently, even boundary dimensions
$d=2r$), we obtain
\begin{eqnarray} 
  \zeta_{\cH}(z) \eq \ln R\,\int_0^\infty
  \frac{\dd\beta}{\Gamma(z)^2}\, \sum_{k=0}^r\, \oint \mu(\bm \alpha)
  \left(\big(\tfrac\b2\big)^2+\big(\tfrac{\a_k}2\big)^2\right)^{z-1}\times
  \nn &&\quad\qquad\times \Bigg[\prod_{\substack{0 \leqslant j
        \leqslant r \\ j \neq
        k}}\frac{\cosh\beta-\cos\alpha_j}{\cos\alpha_k-\cos\alpha_j}\Bigg]
  \chi^{so(2,2r)}_{\cH}(\beta; \vec \alpha_k)\,.
  \label{even zeta}
\end{eqnarray}
where 
\begin{equation}
  \mu(\bm\alpha) := \prod_{n=0}^r \frac{\dd\a_n}{2\pi\,i\,\a_n} \qquad
  \text{and} \qquad \vec \alpha_k :=
  (\alpha_0,\dots,\alpha_{k-1},\alpha_{k+1},\dots, \alpha_r)\,.
\end{equation}
Secondly, for even AdS dimensions $d+1=2r+2$ 
(or equivalently, odd boundary dimensions $d=2r+1$), we obtain
\begin{eqnarray}
  \zeta_{1,\cH}(z) \eq \int_0^\infty
  \frac{\dd\beta\,\beta^{2z-1}}{\Gamma(2z)}
  \sum_{k=0}^r\,\oint\, \mu(\bm \alpha)\,
  \frac{\sinh\frac\b2\,(\cosh\frac\b2)^\frac{1+\e}2
    (\cos\frac{\a_k}2)^\frac{1-\e}2}{2\,(\cosh\b-\cos\a_k)}\, \times
  \nn &&\quad\times \Bigg[\prod_{\substack{0 \leqslant j \leqslant r \\ j \neq k}}
    \frac{\cosh\beta-\cos\alpha_j}{\cos\alpha_k-\cos\alpha_j}\Bigg]
  \chi^{so(2,2r+1)}_{\cH}(\beta; \vec \alpha_k)\,,
  \label{odd zeta}
\end{eqnarray}
where $\e=+1/-1$ for bosonic/fermionic spectrum. The subscript 1 in
the zeta function means that it is the \emph{primary} contribution to
the actual zeta function. The remaining contributions will be
introduced later in this paper but they are irrelevant in all the
cases that we are interested in. Both of the formulae \eqref{even
  zeta} and \eqref{odd zeta} involve complex contour integrals in
$\a_n$ with $n=0,1,\ldots,r$. The contour of each $\a_n$ is a circle
enclosing the origin counter-clockwise with $|\a_0|<\cdots<|\a_r|$.
We would like to stress that the expressions \eqref{even zeta} and
\eqref{odd zeta} allow for a large room for various complex integral
tricks: in the companion paper \cite{part_ii}, we will compute the
zeta function of partially-massless higher-spin gravities in arbitrary
dimensions using complex integrals.

The contour integrals in $\alpha_i$ appearing in the above formulae
\eqref{even zeta} and \eqref{odd zeta} may be evaluated by the residue
theorem, and hence reduce to a linear combination of derivatives in
$\alpha_i$. The expression in terms of the $\alpha_i$ contour integral
is compact and useful in many applications but it might be not
explicit enough in other cases. For instance, if one wants to
implement the formula in a computer program, the other expression in
terms of the $\alpha_i$ derivatives would be more convenient. We
derive the latter expression by adapting the standard tools used to
derive the dimension formula from the Weyl character formula.
Applying the expressions to $d=2,3,4,5,6$,
we provide the explicit form of  the CIRZ in the dimensions
which are  the most relevant for physical applications.

The organization of the paper is as follows. In
\hyperref[sec:zeta_function_AdS]{Section \ref{sec:zeta_function_AdS}},
we start by reviewing the one-loop free energy and zeta function in
AdS$_{d+1}$, then rewrite the latter as an integral of the dimension
formula of an $so(d+2)$ representation.
In \hyperref[sec:CIRZ]{Section \ref{sec:CIRZ}}, we derive the CIRZ
formula in arbitrary dimensions, i.e. we show that the zeta function
of any field in AdS$_{d+1}$ can be written as an integral transform of
its $so(2,d)$ character. 
 In
\hyperref[sec:derivative_expansion]{Section
  \ref{sec:derivative_expansion}}, we spell out an alternative form of
the CIRZ where the previously mentioned contour integrals are replaced
by a linear combination of derivatives of the
character. \hyperref[sec: conclusion]{Section \ref{sec: conclusion}}
contains a brief summary and concluding remarks of the paper. Finally, some definitions and
technical details are presented in
\hyperref[app:character_stuff]{Appendix \ref{app:character_stuff}} and
\ref{sec: app B}.

\section{Zeta functions in AdS}
\label{sec:zeta_function_AdS}
In this section, we shall review the basics of the one-loop free
energy and zeta function, and the integral expression of the zeta
function in AdS$_{d+1}$ obtained in \cite{Camporesi1994}.  After
re-expressing the numerator of the integrand in terms of the dimension
formula of an $so(d+2)$ representation, we shall discuss how the first
derivative of the zeta function can be expressed in a ``spectral
integral'' form.

\subsection{One-loop free energy and zeta function}
The one-loop free energy of a  quantum field 
is given by the logarithm of the one-loop path integral: 
\be
	\Gamma^{\sst (1)}_{[m^2;\cV]}
	=\frac{\e}2\,\log\det_{\cV}(\Box+m^2)\,,
	\label{1l FE}
\ee
where $\cV$ is the space of the off-shell fields
which are traceless and transverse, 
and the sign $\e$ is $+1$ for a boson and $-1$ for a fermion.
The operator $\Box+m^2$ is what appears in the quadratic Lagrangian,
so one can regard the field practically as a free one.
When the field has a gauge symmetry, we have to subtract the 
corresponding ghost contribution.
The one-loop free energy \eqref{1l FE} can be related to the zeta function 
\be
	\zeta_{[m^2;\cV]}(z)=\underset{\cV}{\tr}\left[\frac1{(\Box+m^2)^{z}}\right],
	\label{zeta}
\ee
where the trace is convergent for a sufficiently large value of $\text{Re}(z)$. 
Once we obtain $\zeta_{[m^2;\cV]}(z)$, 
the log det formula can be related to the zeta function 
by analytically continuing the value of $z$ to zero as 
\be
	\log\det_{\cV}(\Box+m^2)
	=\underset{\cV}{\tr}\log(\Box+m^2)
	\ \rightarrow\ 
	-\frac{\e}2\,\zeta'_{[m^2;\cV]}(0),
	\label{zeta-prime}
\ee
where we have used the zeta function regularization and 
the last expression is the finite part of the free energy. 
The UV divergence of the free energy corresponds to $\zeta_{[m^2;\cV]}(0)$.
The zeta function can be also related to the integrated propagator in the
coincidence point limit:
\be
	G_{[m^2;\cV]}=\underset{\cV}{\tr}\left[\frac1{\Box+m^2}\right]
	\ \rightarrow\ 
	\lim_{z\to 1} \zeta_{[m^2;\cV]}(z)\,.
\ee
As we have just seen,  both the free energy
$\Gamma^{\sst (1)}_{[m^2;\cV]}$ and the propagator $G_{[m^2;\cV]}$
can be obtained from  the zeta function \eqref{zeta},
hence this allows us to focus on the zeta function
for a given field space $\cV$ with the mass squared $m^2$.

In AdS$_{d+1}$, free fields can be classified by the irreducible
representations (irreps) they carry for the isometry algebra
$so(2,d)$.  The massive and massless \cite{Metsaev:1995re,
    Metsaev:1997nj, Metsaev:1998xg} irreps\footnote{Notice that there exists an
    ``exotic'' class of field in AdS$_{d+1}$, namely {\it continuous
      spin} fields \cite{Metsaev:2016lhs,
      Metsaev:2017ytk, Metsaev:2017myp}. It was shown that the
    partition function of such fields is equal to one.} are the lowest-weight
modules labeled by $[\Delta;\Y]$, where $\Delta$ is the lowest
eigenvalue of the energy operator generating $so(2)$, and $\Y:=(s_1,
\dots, s_r)$ is the highest weight of the rotational symmetry $so(d)$
classifying the traceless and transverse tensors and can be
interpreted as the spin of the field.  The zeta function for the
module $[\Delta;\Y]$ has been calculated in \cite{Camporesi:1994ga,
  Camporesi1994} (see also e.g.  \cite{Giombi:2016pvg,
  Gunaydin:2016amv} for a review) and its expression is \be
\zeta_{[\Delta;\Y]}(z) = \frac{{\rm Vol}(AdS_{d+1})}{{\rm Vol}(S_d)}\,
\frac{\dim_\Y^{so(d)}}{2^{d-1}\,\G(\frac{d+1}2)^2} \int_0^\infty \dd u
\, \frac{\mu_\Y(u)}{\big[ u^2 + (\Delta - \frac d2)^2 \big]^z}\,.
  \label{zeta_camporesi_higuchi}
\ee
The volume of the $d$-sphere and the (regularized)
volume of the $(d+1)$-dimensional AdS spacetime are given respectively
by
\begin{equation}
  {\rm Vol}(S_d) = \frac{2 \,
    \pi^{\frac{d+1}2}}{\Gamma(\frac{d+1}2)}\, , \qquad {\rm
    Vol}(AdS_{d+1}) = \left\{
  \begin{array}{cc}   
    \frac{2\, (-1)^r\, \pi^{d/2}}{\Gamma(\frac d2 +1)}\,
    \ln R\, \quad & \quad [d = 2r] 
    \medskip \\ 
    \pi^{d/2}\, \Gamma(-\frac d2) \quad & \quad [d = 2r+1]
  \end{array}
  \right.,
\end{equation}
where $R$ is the radius of the AdS$_{d+1}$.
The function $\mu_\Y(u)$ appearing in the numerator of the integrand is given by
\begin{equation}
  \mu_\Y(u) = \prod_{k=1}^r (u^2 + \ell_k^2) \times \left\{
  \begin{array}{cc}
    1 \qquad & \qquad [d=2r]
    \medskip\\
    u \tanh^{\epsilon}(\pi\,u)  \qquad & \qquad [d=2r+1]
  \end{array}
  \right.,
\end{equation}
where $\ell_k=s_k+\frac d2 -k$, and the sign $\epsilon$ is positive for
bosonic fields ($s_k\in\N$) 
and negative for fermionic ones
($s_k\in\frac12 \N$).  The combination $\dim_\Y^{so(d)} \mu_\Y(u)$ in
\eqref{zeta_camporesi_higuchi} is related to the dimension formula
(which is also referred to as Weyl dimension formula) of an $so(d+2)$
irrep:
\begin{itemize}  
\item For even $d=2r$, the dimension of the $so(d+2)$ irrep $(s_0,
  \Y)=(s_0, s_1, \dots, s_r)$ is
  \begin{equation}
    \dim^{so(d+2)}_{(s_0, \Y)} = \prod_{0 \leqslant i < j \leqslant r}
    \frac{(s_i-s_j+j-i)(s_i+s_j+d-i-j)}{(j-i)(d-i-j)}\, .
    \label{dim_so2+d_even}
  \end{equation}
  This can be expressed in terms of the dimension of the $so(d)$ irrep
  $\Y$ as
  \begin{equation}
    \dim^{so(d+2)}_{(s_0, \Y)} = \frac{2\, \dim_\Y^{so(d)}}{d!}\,
    \prod_{k=1}^r (s_0-s_k+k)(s_0+s_k+d-k)\,.
    \label{dim_even}
  \end{equation}
  For $s_0 = i u - \frac d2$, the relation reduces to
  \begin{equation}
    \dim^{so(d+2)}_{(i u - \frac d2, \Y)} = \frac{2\, (-1)^r }{d!}\,
    \dim_\Y^{so(d)}\, \mu_\Y(u)\, .
  \end{equation}
\item For odd $d=2r+1$, the dimension of the $so(d+2)$ irrep $(s_0,
  \Y)$ is
  \begin{eqnarray}
    \dim^{so(d+2)}_{(s_0,\Y)} = \prod_{k=0}^r \frac{2s_k+d-2k}{d-2k}
    \prod_{0 \leqslant i < j \leqslant r}
    \frac{(s_i-s_j+j-i)(s_i+s_j+d-i-j)}{(j-i)(d-i-j)}\,,\,\,
    \label{dim_so2+d_odd}
  \end{eqnarray}
  and can also be related to the dimension of the $so(d)$ irrep $\Y$
  through
  \begin{equation}
    \dim^{so(d+2)}_{(s_0,\Y)} = \frac{\dim^{so(d)}_\Y}{d!}\,
    (2s_0+d)\, \prod_{k=1}^r (s_0-s_k+k)(s_0+s_k+d-k)\, .
  \end{equation}
  For $s_0=iu-\frac d2$, the relation becomes
  \begin{equation}
    \frac{i}2\, \tanh^\epsilon(\pi u)\, \dim^{so(d+2)}_{(iu-\frac
      d2;\Y)} = \frac{(-1)^{r+1}}{d!}\, \dim^{so(d)}_\Y\, \mu_\Y(u)\,.
  \end{equation}
\end{itemize}  
Making use of the above information, the zeta function can be written
as
\be\label{eq:zetachar}
  \zeta_{[\Delta;\Y]}(z) = \int_0^\infty \frac{\dd
    u\,\rho_{\e}(u)}{\big[u^2 + (\Delta-\frac d2)^2 \big]^z}
  \,\dim^{so(d+2)}_{(iu-\frac d2;\Y)} \,,
\ee
with the function $\rho_\e(u)$,
\begin{equation}
  \rho_\e(u)=\left\{
  \begin{aligned}
    & \quad \frac{\ln R}{\pi}\qquad & [{\rm even}\ d]
    \medskip\\
    &\frac{i}2\,\tanh^{\e}(\pi\,u) \qquad &  [{\rm odd}\ d]
  \end{aligned}
  \right.\,.
  \label{def_rho}
\end{equation}
The fact that the zeta function can be written in terms of the
$so(d+2)$ irrep dimension $\dim^{so(d+2)}_{(iu-\frac d2;\Y)}$ helps us
to make a link between the zeta function and the $so(2,d)$
character. Before establishing such a connection, let us first explore
a few interesting properties of the zeta functions.

\subsection{Spectral integral form of the zeta function}
\label{sec:zeta_identities}
The zeta function $\zeta_{[m^2;\cV]}(z)$, defined generally
as \eqref{zeta}, enjoys a simple identity,
\be
	\frac{\partial}{\partial m^2}\zeta_{[m^2;\cV]}(z)
	=\frac1{2m}\,\frac{\partial}{\partial m}\zeta_{[m^2;\cV]}(z)
	=-z\,\zeta_{[m^2;\cV]}(z+1)\,,
	\label{zeta id}
\ee
which is nothing but the spectral version of the Hurwitz zeta function
identity,
\be
	\frac{\partial}{\partial a}\,\zeta(z,a)
	=-z\,\zeta(z+1,a)\,,
\ee
with
\be
\zeta(z,a)=\sum_{n=0}^\infty \frac 1 {(n+a)^z}.
\ee
The identity \eqref{zeta id} simply implies
\be
	\frac{\partial}{\partial m^2}\zeta_{[m^2;\cV]}(0)
	=-\lim_{z\to 0}z\,\zeta_{[m^2;\cV]}(z+1)\,,
\ee
and 
\be
	\frac{\partial}{\partial m^2}\zeta_{[m^2;\cV]}'(0)
	=-\,{\rm F.p.}\lim_{z\to 0}\zeta_{[m^2;\cV]}(z+1)\,,
	\label{zeta rel}
\ee
where F.p. refers to the finite part in the limit $z\rightarrow 0$,
i.e. the constant term in the Laurent expansion in $z$.
These two formulae  provide the derivatives with respect to $m^2$
of the UV divergent and finite part of the free energy.
The second equation \eqref{zeta rel} can be viewed as the regularized version of the formal
expression,
\be
	\frac{\partial}{\partial m^2}\underset{\cV}{\tr}\log(\Box+m^2)=\underset{\cV}{\tr}\left[\frac1{\Box+m^2}\right].
\ee
Now considering the AdS background, the identity \eqref{zeta id}
becomes
\be
	\frac1{2(\D-\frac{d}2)}\,\frac{\partial}{\partial \D}\zeta_{[\D;\Y]}(z)
	=-z\,\zeta_{[\D;\Y]}(z+1)\,,
	\label{zeta id AdS}
\ee
and we also have relations analogous to \eqref{zeta rel}.  These identities
prove useful since it is easier to study the zeta function
$\zeta_{[\D;\Y]}(z)$ near $z=1$ than $z=0$.

In the following, we shall make use of the identity \eqref{zeta id
AdS} to show
how  a ``spectral integral'' form of the
$\zeta'_{[\D;\Y]}(0)$ can be obtained. 
In the context of AdS$_{2r+1}$/CFT$_{2r}$ correspondence, 
this spectral integral formula was used to show the direct relation between 
$\zeta'_{[\D;\Y]}(0)$ and conformal anomaly coefficients. 
It first appeared in the case of totally symmetric representations in \cite{Giombi:2013yva}  
and subsequently mixed symmetry representations in AdS$_7$ \cite{Beccaria:2014xda},
then generalized to arbitrary representations in \cite{Gunaydin:2016amv}.  
Below, we provide a short derivation 
of the spectral integral formula in AdS$_{d+1}$ for both even $d=2r$ and odd $d=2r+1$.

\subsubsection*{Even $d$}
For even values of $d$, the zeta function has the form,
\be
  \zeta_{[\Delta;\Y]}(z) = 
  \int_0^\infty \frac{\dd u\,h_{\Y}(u)}{\big[u^2 + \bar\Delta^2 \big]^z}\,,
  \label{zeta h}
\ee
where $\bar \Delta=\Delta-\frac d2$. 
Since $h_{\Y}(u)=\frac{\ln
  R}{\pi}\,\dim^{so(d+2)}_{(iu-\frac d2;\Y)}$ is an even polynomial of
order $2r$, 
the integral \eqref{zeta h} is the same as 
one half of the integral from $u=-\infty$ to $u=\infty$ with the same integrand.  
The integral is convergent in the region $z>r+\frac12$,
so we can close the contour by adding the infinite upper
half-circle, then shrink it down to enclose the branch cut singularity (i.e. the line defined by ${\rm Arg}(u-i|\bar\Delta|)=\pi$),
\be \zeta_{[\Delta;\Y]}(z) =
\frac12\oint_{i\,|\bar\Delta|} \frac{\dd u\,h_{\Y}(u)}{\big[u^2 +
    \bar\Delta^2 \big]^z}\,.
  \label{contour zeta}
\ee
This contour integral is convergent for any value of $z$, hence we can
directly replace $z$ by the value we want.  If we put $z=0$, the
integrand becomes analytic and we get
\be
	\zeta_{[\Delta;\Y]}(0) =0\,. 
\ee
If we put $z=1$, the integrand has a simple pole at
$u=+i\,|\bar\Delta|$ and gives
\be
	\zeta_{[\Delta;\Y]}(1)=\frac12\,2\pi i\,\frac{h_\Y(i\,|\bar\Delta|)}{2i\,|\bar \Delta|}
	=\pi\frac{h_\Y(i\,\bar\Delta)}{2\,|\bar \Delta|}\,.
\ee
Using \eqref{zeta id AdS}, this implies
\be
	\frac{\partial}{\partial \D}\zeta_{[\D;\Y]}'(0)
	=-2\,\bar\Delta\,\zeta_{[\D;\Y]}(1)
	=-\pi\,{\rm sgn}(\bar\Delta)\,h_\Y(i\,\bar\Delta)\,.
	\label{zeta' id}
\ee
From the fact that $\zeta_{[\frac d2;\Y]}'(0)=0$ we can derive the
expression
\be
	\zeta_{[\D;\Y]}'(0)
	=-\pi\,\int^{|\bar\Delta|}_0 \dd x\,h_\Y(i\,x)
	=-\ln R\,
	\int^{|\bar\Delta|}_0 \dd x\,\dim^{so(d+2)}_{(-x-\frac d2;\Y)}\,,
	\label{zeta' int}
\ee
where the absolute value $|\bar\Delta|$ appears as a result of
$\rm sgn(\bar \Delta)$.  The result is even in $\bar\Delta$ like the
original form \eqref{zeta h}
hence insensitive to its sign.
In physical term, the sign of $\bar\Delta$ determines
whether the underlying field takes Dirichlet or Neumann boundary condition.
Since the two boundary conditions should give 
different results, we need to modify the above definition of the zeta function.

By noticing that the expression \eqref{zeta' int} is not analytic 
on the imaginary axis of $\bar\Delta$,
we can consider another expression where we analytically continue the value of $\bar\Delta$
from positive ${\rm Re}(\bar\Delta)$ to negative one.
This simply amounts to replacing
$|\bar\Delta|$ by $\bar\Delta$ in \eqref{zeta' int}. 
At the level of the contour integral
representation \eqref{contour zeta}, this ``new'' definition of the zeta function corresponds to
the modification,
\be
  \zeta_{[\Delta;\Y]}(z) = 
  \frac12\oint_{i\,\bar\Delta} \frac{\dd u\,h_{\Y}(u)}{\big[u^2 + \bar\Delta^2 \big]^z}\,,
  \label{zeta-analytic}
\ee
where the contour encircles counter-clockwise the branch cut starting
at $i\,\bar \Delta$ rather than $i\,|\bar\Delta|$\,. 
The zeta function \eqref{zeta-analytic} is what has been used in the literature. 

\subsubsection*{Odd $d$}
For odd values of $d$, the zeta function has the form,
\be
  \zeta_{[\Delta;\Y]}(z) = 
  \int_0^\infty \frac{\dd u\,\tanh^\e(\pi u)\,h_{\Y}(u)}{\big[u^2 + \bar\Delta^2 \big]^z}\,,
  \label{zeta h e}
\ee
where $h_{\Y}(u)=\frac i2\,\dim^{so(d+2)}_{(i\,u-\frac d2;\Y)}$ is now an odd function.
Similarly to the even $d$ case, we can rewrite the above expression as a contour integral 
 by adding to the real line the infinite radius upper-half circle.
The function $h_{\Y}(u)$ is analytic again but $\tanh^\e(\pi\,u)$ has infinitely many 
simple poles on the imaginary axis. We can separate those contributions as
\be
  \zeta_{[\Delta;\Y]}(z) = 
  \frac12\left(\oint_{i\,\bar\Delta} 
  +\sum_{n=1}^\infty \oint_{i\left(n-\frac{1+\e}4\right)}\right)
  \frac{\dd u\,\tanh^\e(\pi u)\,h_{\Y}(u)}{\big[u^2 + \bar\Delta^2 \big]^z}\,.
  \label{zeta h e 2}
\ee
Here, we take the prescription that the zeta function is analytic in $\bar \Delta$.
Due to the presence of  infinitely many simple poles of $\tanh^\e(\pi u)$,
it is not easy to simplify further the above expression as opposed to the even $d$ case.
However, if we take the difference between  $\zeta_{[\Delta;\Y]}(z)$
and $\zeta_{[d-\Delta;\Y]}(z)$, we can cancel the cumbersome contribution 
and end up with
\be
	\zeta_{[\Delta;\Y]}(z)-\zeta_{[d-\Delta;\Y]}(z)
	=\oint_{i\,\bar\Delta} 
  \frac{\dd u\,\tanh^\e(\pi u)\,h_{\Y}(u)}{\big[u^2 + \bar\Delta^2 \big]^z}\,.
  \label{z integ}
\ee
Note that we do not have a perfect cancellation 
due to the prescription of analytic continuation in $\Delta$. 
If $\Delta$ is not an integer/half-integer for boson/fermion,
taking $z=0$ and $z=1$ limit, we find
\be
	\zeta_{[\Delta;\Y]}(0)-\zeta_{[d-\Delta;\Y]}(0)=0\,,
	\label{even zeta 0 diff}
\ee
and
\be
	\zeta_{[\Delta;\Y]}(1)-\zeta_{[d-\Delta;\Y]}(1)=\pi\,\frac{\tanh^\e(\pi\,i\,\bar\Delta)\,h_\Y(i\,\bar\Delta)}{\bar\Delta}\,.
	\label{even zeta 1 diff}
\ee
The first equation \eqref{even zeta 0 diff} means that the UV divergence 
does not depend on the sign of $\bar\Delta$, or in physical terms, the choice of the boundary conditions.
Applying the second equation  \eqref{even zeta 1 diff} to the zeta function identity \eqref{zeta id AdS}, we reach the result,
\begin{eqnarray}
  \zeta_{[\D;\Y]}'(0) - \zeta_{[d-\D;\Y]}'(0) \eq -2\pi\,
  \int^{\bar\Delta}_0 \dd x\, \tanh^\e(\pi\,i\,x)\, h_\Y(i\,x)\nn \eq
  \e\,\pi \int^{\bar\Delta}_0 \dd x\, \tan^\e(\pi\,x)\,
  \dim^{so(d+2)}_{(-x-\frac d2;\Y)}\,.
  \label{even AdS diff}
\end{eqnarray}
Hence, the free energy difference between $\D$ and $d-\D$ 
is given by a ``spectral integral'' where the integrand 
involves the dimension of an $so(d+2)$ irrep.

If $\Delta$ is an integer/half-integer 
(or equivalently $\bar\Delta$ is an half-integer/integer) for boson/fermion, 
the difference of the zeta zero \eqref{even zeta 0 diff} does not vanish anymore but gives
\be
	\zeta_{[\Delta;\Y]}(0)-\zeta_{[d-\Delta;\Y]}(0)=2\,i\,h_\Y(i\,\bar\Delta)
	=-\dim^{so(d+2)}_{(-\Delta;\Y)}\,.
	\label{even zeta 0 diff'}
\ee 
Moreover, the equations \eqref{even zeta 1 diff} 
should be also modified because the integral \eqref{z integ} with $z=1$ now involves a double pole,
and consequently \eqref{even AdS diff} should be modified as well.
After all, the necessary modification in \eqref{even zeta 1 diff} and \eqref{even AdS diff}
for (half-)integer $\Delta$
amounts to removing the singularity
arising in the limit where $\Delta$ approaches to an integer or half-integer.
The exceptionality of the (half-)integer $\Delta$ was first observed in \cite{Giombi:2013yva},
where the focus was on the massless case with $\Delta=s+d-2$.

Let us delve a little further in the consequences of \eqref{even zeta 0 diff'}.
Let  $w_I$ and $h_I$ denote the number of columns and rows contained in the $I$-th block\,\footnote{In other words, the $I$-th block of $\Y$ is a succession of rows with the same length, and a diagram can be described as an aggregate of blocks ordered by decreasing length when examined from top to bottom (i.e. the first block is the one at the top of the diagram). In the notation introduced above, the $I$-th block is of length $w_I$ and height $h_I$, meaning it is composed of $h_I$ rows which are all of length $w_I$.} of $\Y$,
 and define  $p_I=h_1+h_2+\cdots +h_I$ with $p_0=0$.
 Then,  $w_I=s_{p_{I-1}+1}=\ldots=s_{p_{I}}$
 and  $\Y$ can be denoted by $\Y=(w_1^{h_1},w_{2}^{h_2},\ldots)$.
Now, if we assume that $\Delta$ is an (half-)integer satisfying
\be
	w_{I+1}-p_I\le \Delta-d \le w_{I}-p_I-1\,,
\label{interval}
\ee
then we can use the identity
\be
	 \dim^{so(d+2)}_{(-\Delta,w_1^{h_1},w_{2}^{h_2},\ldots)}
	 =(-1)^{p_I+1}\, \dim^{so(d+2)}_{({(w_1-1)}^{h_1},\ldots,{(w_I-1)}^{h_I}, \Delta-d+p_I,w_{I+1}^{h_{I+1}},\ldots)}\,.
\ee
Note that  the coefficient $\dim^{so(d+2)}_{(-\Delta;\Y)}$ appearing in \eqref{even zeta 0 diff'} does not vanish for a generic (half-) integer 
except for the points $\Delta=s_k+d-k$ with $k=1,\ldots,r$. 
In particular, the massive fields with $\Delta\ge s_1+d$ give the result  $\dim^{so(d+2)}_{(\Delta-d;\Y)}$. 
This AdS result could be reproduced in the CFT side from the zero modes of the effective kinetic operator
of the Hubbard-Stratonovich field.
In \cite{Giombi:2013yva}, the eigenvalues of such an operator in 3d has been calculated for $s=0,1,2$ and conjectured for arbitrary integer spins as
(the equation (3.24) of \cite{Giombi:2013yva})
\be
	k_{n,0}=c_s(\Delta)\,\frac{\Gamma(n-1+\Delta)}{\Gamma(n+2-\Delta)}\,,
	\qquad
	k_{n,i}=\frac{\Gamma(2-\Delta)}{\Gamma(\Delta-1)}\,\frac{\Gamma(-1+i+\Delta)}{\Gamma(2+i-\Delta)}\,k_{n,0}\,,
\ee
where $n$ and $i$ range from $s+1$ to infinity and $-s$ to $s$, respectively. 
The eigenvalue $k_{n,i}$ vanishes if $n\le \Delta-2$ and the degeneracy for a fixed $n$ and $i$ is $n^2-i^2$\,.
Hence the total number of zero modes is
\be
	\sum_{n=s+1}^{\Delta-2} \sum_{i=-s}^{s}\,(n^2-i^2)=\frac{(2s+1)\,(2\,\Delta-3)\,(\Delta-s-2)\,(\Delta+s-1)}{3!}
	=\dim^{so(5)}_{(\Delta-3,s)}\,.
\ee
Indeed, one can see that the number of the zero modes coincides with the AdS result \eqref{even zeta 0 diff'}.

Considering now the (mixed-)symmetric (partially-)massless fields with $\Delta_{\rm\sst PM}=w_I+d-p_I-t$ with $1 \leqslant t \leqslant w_I-w_{I+1}$, i.e. $\Delta_{\rm\sst PM}$ satisfies \eqref{interval},
we obtain 
\be
\zeta_{[\Delta_{\rm\sst PM};\Y]}(0)-\zeta_{[d-\Delta_{\rm\sst PM};\Y]}(0)
	=(-1)^{p_I}\,\dim^{so(d+2)}_{\Y_{\rm\sst KT}}\,,
	\label{phys KT}
\ee
where $\Y_{\rm\sst KT}$ is the $so(d+2)$ irrep carried by the associated Killing tensors:
\be
	\Y_{\rm\sst KT}=({(w_1-1)}^{h_1},\ldots, {(w_I-1)}^{h_I}, w_I-t,w_{I+1}^{h_{I+1}},\ldots)\,.
\ee 
Considering the gauge parameter of the same field 
having $\Delta_{\rm\sst GP}=w_I+d-p_I$ and $\Y_{\rm\sst GP}=(w_1^{h_1}, \ldots,w_{I-1}^{h_{I-1}}, {w_I}^{h_I-1}, w_I-t,w_{I+1}^{h_{I+1}},\ldots)$,
we find again the dimension of the Killing tensors:
\be
\zeta_{[\Delta_{\rm\sst GP};\Y_{\rm\sst GP}]}(0)-\zeta_{[d-\Delta_{\rm\sst GP};\Y_{\rm\sst GP}]}(0)
	=(-1)^{p_I+1}\,\dim^{so(d+2)}_{\Y_{\rm\sst KT}}\,,
	\label{gauge KT}
\ee
but with opposite sign. 
Since the full zeta function is the difference between the physical mode and the gauge mode contributions, 
the net result becomes two times of \eqref{phys KT}.
Like in the massive integral $\Delta$ case, the above AdS result for (partially-)massless field
could be reproduced from the zero modes of the effective CFT kinetic operators.
On top of these, the contributions of the ghost zero modes, giving rise again to the dimension of Killing tensors, should be appended to 
both sides of AdS and CFT,
and it was shown in \cite{Giombi:2013yva} that they match each other as a  consequence of `AdS Killing tensor = Conformal Killing tensor'.

\section{Contour integral expression of the CIRZ}
\label{sec:CIRZ}

Now we turn to the main objective of the current paper --- the
derivation of the character integral representation of the zeta
function in any dimensional AdS spacetime.  Our goal is to express the
zeta function \eqref{eq:zetachar} in terms of the $so(2,d)$ character
so that the dependence on $\Delta$ and $\Y$ in $\zeta_{[\Delta;\Y]}$
enters only through the corresponding character
$\chi_{(\Delta,\Y)}^{so(2,d)}$ (see e.g. \cite{Dolan:2005wy,
  Beccaria:2014jxaw, Basile:2016aen, Bourget:2017kik} for more details
on $so(2,d)$ characters).  It turns out that it is sufficient to
consider the character solely over (possibly reducible) generalized
Verma modules of the conformal algebra. These representations, when
irreducible, correspond to massive fields in AdS for which the
one-loop partition function takes the form \eqref{1l FE}. The
formalism we develop extends trivially to massless fields
\cite{Bae:2016hfy, Gupta:2012he}, which are described by irreducible
representation defined as quotients of generalized Verma modules.

\subsection{General dimensions}
The character of the $so(2,d)$ generalized Verma module ${\cal
  V}(\Delta;\Y)$ takes the form,
\begin{equation}
  \chi^{so(2,d)}_{(\Delta;\Y)}(\beta;\vec \alpha) = e^{-\beta\Delta}\,
  \chi^{so(d)}_{\Y}(\vec \alpha)\, \Pd d (i\beta;\vec \alpha)\,,
  \label{def_character_so2d}
\end{equation}
where $\Pd d$ is defined as
\begin{equation}
  \Pd d (\alpha_0; \vec \alpha) = \frac{e^{-i\,\frac d2\,\alpha_0}}{2^{d-r}}\,
  \prod_{k=1}^r
  \frac{1}{\cos\alpha_0-\cos\alpha_k}
  \times
  \left\{
  \begin{aligned}
  &\quad 1 & \qquad [{\rm even}\ d]\\
  & \frac i{\sin\frac{\a_0}2}
  & \qquad [{\rm odd}\ d]\\
  \end{aligned}\right.,
  \label{def_Pd}
\end{equation}
with $r=[d/2]$ and $\vec \a=(\a_1,\ldots,\a_r)$. We first note that  the dimension of the $so(d+2)$ irrep  
is given by the corresponding character 
evaluated at $\bm\a=\bm 0$:
\be
	 \dim^{so(d+2)}_{(iu-\frac
    d2;\Y)}
    =\left[\chi^{so(d+2)}_{(iu-\frac d2;\Y)}(\bm \alpha)\right]_{\bm
      \alpha = \bm 0}\,,
    \label{Weyl_dim_formula}
\ee
where $\bm \alpha = (\alpha_0, \dots, \alpha_r)$ and $r=[d/2]$.  Then,
the $so(d+2)$ character can be related to that of $so(d)$ using the
following identity (see \hyperref[app:character_stuff]{Appendix
  \ref{app:character_stuff}} for additional details on the identity):
\begin{equation}
  \chi^{so(d+2)}_{(s_0,\Y)}(\bm \alpha) = \sum_{k=0}^r \left(\,
  e^{-i\alpha_k\,s_0}\, \chi^{so(d)}_{\Y_-}(\vec \alpha_k) + (-1)^d\,
  e^{i\alpha_k\, (s_0+d)}\, \chi^{so(d)}_{\Y_+}(\vec
  \alpha_k)\,\right) \Pd d (\alpha_k;\vec \alpha_k)\,,
  \label{prop_character_so2+d}
\end{equation}
where $\vec\alpha_k=(\a_0,\ldots, \a_{k-1}, \a_{k+1},\ldots,\a_r)$ and
$\Y_\pm =(s_1, \dots, s_{r-1}, (\pm)^{d+1} s_r)$\,.  Another key trick
is based on the identity,
\begin{equation}
  \frac{1}{\big[ u^2 + \bar\Delta^2 \big]^z} =
  \frac{\sqrt{\pi}}{\Gamma(z)} \, \int_0^\infty \dd \beta \left(
  \frac{\beta}{2u} \right)^{z-\frac12}\, J_{z-\frac12}(\beta\,u)\,
  e^{-\beta\,\bar\Delta}\, ,
  \label{prop_integral_beta}
\end{equation}
holding for ${\rm Re}(\bar\Delta)>0$ and ${\rm Re}(z) > 0$.  Note here
that for the convergence of the $\beta$ integral, we ought to use
$e^{-\beta\,|\bar\Delta|}$ in the $\beta$ integral, but in such a case
the zeta function will be insensitive to the sign of $\bar\Delta$ and
becomes incapable of distinguishing different boundary conditions.
Hence, like the discussion below \eqref{zeta' int}, we first derive
the formula assuming ${\rm Re}(\bar \Delta)>0$ then analytically
continue $\bar\Delta$ to the negative ${\rm Re}(\bar \Delta)$ region.
Combining all these elements, we get
\begin{eqnarray}
  && \zeta_{[\Delta;\Y]}(z) = \frac{\sqrt{\pi}}{\Gamma(z)}
  \int_0^\infty \dd u \int_0^\infty \dd \beta \, \rho_\e(u)\,
  \left(\frac{\beta}{2u} \right)^{z-\frac12}\, J_{z-\frac12}(\beta\,u)
  \,\\ && \quad \times \bigg[\sum_{k=0}^r e^{\frac d2
      (\beta+i\alpha_k)}\, \frac{\Pd d (\alpha_k;\vec \alpha_k)}{\Pd d
      (i\beta;\vec \alpha_k)}\, \Big( e^{u\alpha_k}\,
    \chi^{so(2,d)}_{(\Delta ;\Y_+)}(\beta; \vec \alpha_k) +
    (-1)^d\,e^{-u\alpha_k}\, \chi^{so(2,d)}_{(\Delta ;\Y_-)}(\beta;
    \vec \alpha_k) \Big) \bigg]_{\bm \alpha = \bm 0}\, ,\nonumber
  \label{interm}
\end{eqnarray}
where we used the factor $e^{-\beta(\Delta-\frac d2)}$ to reconstruct
the $so(2,d)$ characters of the irreps $[\Delta;\Y_\pm]$ according to
\eqref{def_character_so2d}.  We can simplify the above formula using
the identity of the $so(d)$ characters,
\begin{equation}
  \chi^{so(d)}_{\Y_-}(\alpha_1, \dots, \alpha_{k-1}, -\alpha_k, \alpha_{k+1}, \dots, \alpha_{r-1}, \alpha_r) =
  \chi^{so(d)}_{\Y_+}(\alpha_1, \dots, \alpha_{k-1}, \alpha_k, \alpha_{k+1}, \dots, \alpha_r)\,, 
\end{equation}
for $k=1,\dots,r$. We obtain
\begin{eqnarray}
  && \zeta_{[\Delta;\Y]}(z) = \frac{\sqrt{\pi}}{\Gamma(z)}
  \int_0^\infty \dd u \int_0^\infty \dd \beta \left(\frac{\beta}{2u}
  \right)^{z-\frac12}\, J_{z-\frac12}(\beta\,u) \,\nn && \qquad \times
  \bigg[ \sum_{k=0}^r \tilde\nu_\e(u,\b,\a_k) \,\bigg(\frac{\sinh\frac\b2}{\sin\frac{\a_k}2}\bigg)^{d-2r}\!
 \prod_{\substack{0
        \leqslant j \leqslant r \\ j \neq k}}    \frac{\cosh\beta-\cos\alpha_j}{\cos\alpha_k-\cos\alpha_j}\,
    \chi^{so(2,d)}_{(\Delta ;\Y)}(\beta; \vec \alpha_k) \bigg]_{\bm
    \alpha = \bm 0}\, ,
  \label{zeta_even}
\end{eqnarray}
where the function $\tilde\nu_\e(u,\b,\a)$ is defined to be
\be
	\tilde\nu_\e(u,\b,\a)
	=\left\{
	\begin{aligned}
	&\qquad \frac{2\,\ln R}{\pi}\,\cosh(\a\,u) \qquad & [{\rm
              even}\ d]\\ &-\tanh^{\e}(\pi\,u)\,\sinh(\a\,u)
          \qquad & [{\rm odd}\ d]
	\end{aligned}\right. .
	\label{nu-tilde}
\ee
Note here that each summand in the second line of \eqref{zeta_even}
diverges in the limit $\bm\a\to \bm 0$. Only the sum of the $r+1$ terms 
is regular in the limit $\bm\a\to \bm 0$.  For this
reason, we cannot exchange the order of the summation and the
evaluation $\bm\a=\bm 0$.

To further simplify the formula, it is
convenient to replace the evaluation of the character at $\bm \alpha =
\bm 0$ by the contour integrals,
\begin{equation}
  \dim_{(s_0,\Y)}^{so(d+2)} = \prod_{k=0}^r \oint_{\mathscr{C}_k}
  \frac{\dd \alpha_k}{2\,\pi\,i\, \alpha_k}\,
  \chi^{so(d+2)}_{(s_0,\Y)}(\bm \alpha)\, ,
  \label{dim char}
\end{equation}
where $\mathscr{C}_k$ are contours encircling the origin
counter-clockwise such that the contour $\mathscr{C}_k$ lies inside of
$\mathscr{C}_{k+1}$: e.g. the circular contours with
$|\a_k|<|\a_{k+1}|$ (see \hyperref[fig:contour]{Figure
  \ref{fig:contour}}).
\begin{figure}[!ht]
  \label{fig:contour}
  \center
  \begin{tikzpicture}
    \draw[help lines, color=gray!30, dashed] (-2.4,-2.4) grid (2.4,2.4);
    \draw[->,ultra thick] (-2.5,0)--(2.5,0);
    \draw[->,ultra thick] (0,-2.5)--(0,2.5);
    \draw[thick,color=theRed] (0,0) circle (1cm);
    \node (second) at (-0.5,-0.5) {$\mathscr{C}_0$};
    \draw [thick] (0,0) circle (0.5cm);
    \node[color=theRed] (second) at (0.9,-0.9) {$\mathscr{C}_1$};
    \draw[thick,color=myblue] (0,0) circle (1.5cm);
    \node[color=myblue] (second) at (1.2,1.2) {$\mathscr{C}_2$};
    \draw[->, thick, color=myblue] (-1.06,1.06) arc[radius=1.2cm,start angle=135,delta
      angle=15];
    \draw[->, thick, color=theRed] (0.71,0.71) arc[radius=0.9cm,start angle=45,delta
      angle=12];
    \draw[->, thick] (0.35,-0.35) arc[radius=0.7cm,start angle=310,delta
      angle=12];
  \end{tikzpicture}
  \caption{Example of ``ordered'' contours with $r=2$. }
\end{figure}
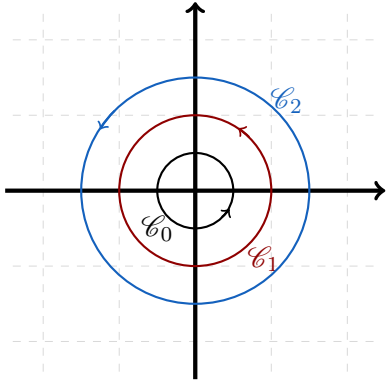

\noindent
The advantage of the contour integral representation is that now we
can perform the contour integration before the summation --- but keeping
the order of contours fixed.  In this way we get
\ba
 &&\zeta_{[\Delta;\Y]}(z) = \frac{\sqrt{\pi}}{\Gamma(z)}
\int_0^\infty \dd \beta \int_0^\infty \dd u  
   \left(\frac{\beta}{2u} \right)^{\!z-\frac12}\, J_{z-\frac12}(\beta\,u) \nn
&& \ \times\,
  \sum_{k=0}^r\oint_C\mu(\bm \a)\,\tilde\nu_\e(u,\b,\a_k)\,\bigg(\frac{\sinh\frac\b2}{\sin\frac{\a_k}2}\bigg)^{d-2r}\!\!
  \prod_{\substack{0 \leqslant j \leqslant r \\ j \neq
        k}}
    \frac{\cosh\beta-\cos\alpha_j}{\cos\alpha_k-\cos\alpha_j}\,
 \chi^{so(2,d)}_{(\Delta ;\Y)}(\beta; \vec
    \alpha_k)\,,\quad   \label{zeta_even'}
\ea
where we have interchanged the order of $u$ and $\b$ integrations
and  $\mu(\bm \a)$ denotes the measure for $(r+1)$-variable complex integral,
\be
	\oint_C\mu(\bm \a)=\prod_{k=0}^r \,\underset{|\a_k|<|\a_{k+1}|}{\oint}\,
  \frac{\dd \alpha_k}{2\,\pi\,i\, \alpha_k}\,.
\ee
Finally, exchanging the order of the $u$ and $\bm\a$ integrations, we obtain
\ba
 \zeta_{[\Delta;\Y]}(z) \eq \frac{\sqrt{\pi}}{\Gamma(z)}\int_0^\infty \dd \beta 
 \sum_{k=0}^r\oint_C \mu(\bm \a)\,
 \nu_\e(z,\b,\a_k)\times \nn
 &&\quad \times\,
 \bigg(\frac{\sinh\frac\b2}{\sin\frac{\a_k}2}\bigg)^{d-2r}\!
  \prod_{\substack{0 \leqslant j \leqslant r \\ j \neq
        k}}
    \frac{\cosh\beta-\cos\alpha_j}{\cos\alpha_k-\cos\alpha_j}\,
 \chi^{so(2,d)}_{(\Delta ;\Y)}(\beta; \vec
    \alpha_k)\,,  \label{zeta intmed}
\ea
where the function $\nu_\e(z,\b,\a)$ 
is defined to be
\ba\label{nu}
	&&\nu_\e(z,\b,\a)=\int_0^\infty \dd u \,  \tilde\nu_\e(u,\b,\a)\,
   \left(\frac{\beta}{2u} \right)^{\!z-\frac12}\, J_{z-\frac12}(\beta\,u)\nn
   &&=
   \int_0^\infty \dd u \,
   \left(\frac{\beta}{2u} \right)^{\!z-\frac12}\, J_{z-\frac12}(\beta\,u)\,
   \times\left\{
	\begin{aligned}
	&\qquad \frac{2\,\ln R}{\pi}\,\cosh(\a\,u) \quad & [{\rm
              even}\ d]\\ &-\tanh^{\e}(\pi\,u)\,\sinh(\a\,u)
          \quad & [{\rm odd}\ d]
	\end{aligned}\right. .
\ea
Evaluation of the above function requires separate considerations for
even and odd $d$.

\subsection{AdS$_{2r+1}$}
\label{subsec:zeta_character_even}
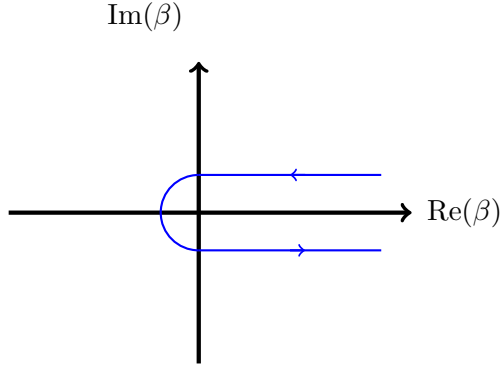
\begin{figure}[!ht]
  \label{fig:contour}
  \center
  \begin{tikzpicture}
    \draw[->,ultra thick] (-2.5,0)--(2.8,0); \node (xaxis) at (3.5,0) {${\rm Re}(\beta)$};
    \draw[->,ultra thick] (0,-2)--(0,2); \node (yaxis) at (-0.7,2.6) {${\rm Im}(\beta)$};
    \draw[thick,blue] (0,0.5) arc[radius=0.5cm,start angle=90,delta
      angle=180];
     \draw[thick,blue] (2.4,0.5)--(0,0.5); \draw[thick,blue] (0,-0.5)--(2.4,-0.5);
     \draw[->,thick,blue] (1.4,0.5)--(1.2,0.5); \draw[->,thick,blue] (1.2,-0.5)--(1.4,-0.5);
  \end{tikzpicture}
  \caption{Hankel contour.}
  \label{Hankel contour}
\end{figure}

In even $d=2r$, the function $\nu_\e(z,\b,\a)$ can be evaluated as
\be\label{nu even d}
\nu_\e(z,\b,\a)= \frac{2\,\ln R}{\pi}\int_0^\infty \dd u \,
\cosh(\alpha\,u)\, \left(\frac{\beta}{2u} \right)^{\!z-\frac12}\,
J_{z-\frac12}(\beta\,u)\ =\frac{\ln
  R}{\sqrt{\pi}}\,\frac{\left(\big(\frac\b2\big)^2+\big(\frac\a2\big)^2\right)^{z-1}}{\Gamma(z)}\,.
\ee
This integral was evaluated in the region 
${\rm Re}(\a)=\rm{Im}(\b)=0,\,\b>0$ and ${\rm Re}(z)>0$,
then analytically continued to other region.
Since an additional factor of $1/\G(z)$ is generated, we can express
the result as
\be
	 \G(z)\,\zeta_{[\Delta;\Y]}(z) = \ln R\, \int_0^\infty 
	 \frac{\dd \beta}{\G(z)}\,\left(\frac\b 2\right)^{2(z-1)}\,
	 f_{[\Delta;\Y]}(z,\b),
	 \label{GZf}
\ee
where the function $f_{[\Delta;\Y]}(z,\b)$ is defined as
\be
	f_{[\Delta;\Y]}(z,\b)
	=\sum_{k=0}^r\oint_C \mu(\bm \a)\,
	\left(1+\big(\tfrac{\a_k}{\b}\big)^2\right)^{z-1}\,
  \prod_{\substack{0 \leqslant j \leqslant r \\ j \neq
        k}}
    \frac{\cosh\beta-\cos\alpha_j}{\cos\alpha_k-\cos\alpha_j}\,
 \chi^{so(2,d)}_{(\Delta ;\Y)}(\beta; \vec
    \alpha_k)\,.
    \label{f even}
\ee
Note that the contour of $\a_k$ does not contain $\pm i\,\b$.  The
function $\b^{2(z-1)}$ has a branch cut on the negative real
axis of $\b$ with the phase factor $e^{4\pi z i}$. Using this
information, we can rewrite \eqref{GZf} as
\be
	 \Gamma(z)\,\zeta_{[\Delta;\Y]}(z) = \ln R\,  \oint \frac{\dd \beta}{\Gamma(z)\,2\,i\,\sin(2\pi z)}\,
	\,\left(-\frac\b 2\right)^{2(z-1)}\, f_{[\Delta;\Y]}(z,\b)\,,
	 \label{zeta z even}
\ee
where the contour integral is defined by the Hankel contour  (see Figure \ref{Hankel contour}).

\noindent Therefore, the first derivative of the zeta function  reduces to
\begin{equation}
  \zeta'_{[\Delta;\Y]}(0) = \frac{\ln R}{2}\, \oint
  \frac{\dd\beta}{2\pi\,i}\,\left(\frac \b {2}\right)^{\!-2}\,
  f_{[\Delta;\Y]}(0,\b)\,,
  \label{zeta'}
\end{equation}
whereas $ \zeta_{[\Delta;\Y]}(0) =0$. Notice that, because the
integrand of the above integral is devoid of branch cut, the Hankel
contour can be deformed to a closed contour encircling the origin of
the complex plane.  In the end, we arrived at the relation between the
zeta function $\zeta_{[\Delta;\Y]}(z)$ and the character
$\chi^{so(2,d)}_{(\Delta;\Y)}(\beta; \vec \alpha)$ for an arbitrary
lowest weight representation $[\Delta;\Y]$\,.  Since the relation
\eqref{zeta'} is a linear map independent of the representation, we
can apply the same formula to a generic 
\emph{reducible}
representation of the $so(2,d)$ algebra,
e.g. the representation corresponding to the field content of any theory.

\subsection{AdS$_{2r+2}$}
\label{sec: odd d CIRZ}
In odd $d=2r+1$, the function $ \nu_\e(z,\b,\a)$ becomes 
\begin{equation}
  \nu_\e(z,\b,\a)= - \frac{\big(\frac
    \b2\big)^{2z-1}}{\Gamma(z+\frac12)}
  \,\l_\e(z,\b,\a),
 \label{nu lambda}
\end{equation}
 with
 \be
 \l_\e(z,\b,\a)=
 \int_0^\infty \dd u \,  \tanh^\e(\pi\,u)\,\sinh(\alpha\,u)\, {}_0F_1\big(z+\tfrac12,-\tfrac{u^2\,\b^2}{4}\big)\,.
 \label{lambda-epsilon-z}
 \ee
Here, we have used the identity,
 \be
	 \left(\frac{\beta}{2u} \right)^{\!z-\frac12}\, J_{z-\frac12}(u\,\beta)
	 =\frac{\big(\frac \b2\big)^{2z-1}}{\Gamma(z+\frac12)}\,
	 {}_0F_1\big(z+\tfrac12,-\tfrac{u^2\,\b^2}{4}\big)\,.
	 \label{bessel_hypergeo}
\ee
In terms of the function $\l_\e(z,\b,\a)$, the zeta function is given by
\ba
 \zeta_{[\Delta;\Y]}(z) \eq -\int_0^\infty \dd \beta\,\frac{\b^{2z-1}}{\Gamma(2z)}
 \sum_{k=0}^r\oint_C \mu(\bm \a)\,
  \l_\e(z,\b,\a_k)\,
\frac{\sinh\frac\b2}{\sin\frac{\a_k}2}\times\nn
 &&\times\,
  \prod_{\substack{0 \leqslant j \leqslant r \\ j \neq
        k}}
    \frac{\cosh\beta-\cos\alpha_j}{\cos\alpha_k-\cos\alpha_j}\,
 \chi^{so(2,d)}_{(\Delta ;\Y)}(\beta; \vec
    \alpha_k)\,.  \label{zeta lambda}
\ea
If we expand $\l_\e(z,\b,\a)$ in $z$ as \be
	\l_\e(z,\b,\a)
	=\l_\e(0,\b,\a)+z\,\l_\e'(0,\b,\a)+\cO(z^2)\,,
	\label{lambda exp}
\ee
where $\l'_\e(z,\b,\a)=\frac{\partial}{\partial z}\l_\e(z,\b,\a)$\,, the 
higher order terms $\cO(z^2)$ contribute neither to
$\zeta_{[\Delta;\Y]}(0)$ nor to $\zeta'_{[\Delta;\Y]}(0)$ --- the
parts of the zeta function that are relevant to physics.  
The second term $z\,\l'_\e(0,\b,\a)$ does not contribute to
$\zeta_{[\Delta;\Y]}(0)$, but it does to $\zeta'_{[\Delta;\Y]}(0)$.
However, as we shall explain below, its contribution vanishes if the
character is even in $\b$. Moreover, even when they contribute, their
contribution is a rational number.  For this reason, their role is
rather marginal compared to the term $\l_\e(0,\b,\a)$\,.  The zeta
function obtained only with $\l_\e(0,\b,\a)$ neglecting the $\cO(z)$
term will be referred to as the \emph{primary} contribution
$\zeta_{1,[\Delta;\Y]}(z)$,\footnote{The primary contribution
  $\zeta_{1,[\Delta;\Y]}(z)$ was referred to as the modified zeta
  function $\tilde\zeta_{[\Delta;\Y]}(z)$ in \cite{Bae:2016rgm}.}
while the contribution related to $z\,\l'_\e(0,\b,\a)$ as the
\emph{secondary} one $\zeta_{2,[\Delta;\Y]}(z)$.  To summarize, the
physically relevant parts of the zeta function are given by
\be
	\zeta_{[\Delta;\Y]}(0)=\zeta_{1,[\Delta;\Y]}(0)\,,
	\qquad
	\zeta'_{[\Delta;\Y]}(0)=\zeta'_{1,[\Delta;\Y]}(0)+\zeta'_{2,[\Delta;\Y]}(0)\,,
\ee
where $\zeta_{1,[\Delta;\Y]}(z)$ and
$\zeta_{2,[\Delta;\Y]}(z)$\, are calculated hereafter.

\subsubsection*{Primary contribution}

Let us consider first the primary contribution to the zeta function.
It is given through $\l_\e(0,\b,\a)$ in the form of an integral,
\be
	\l_\e(0,\b,\a)
	=\int_0^\infty \dd u \,  \tanh^\e(\pi\,u)\,\sinh(\alpha\,u)\,\cos(\beta\, u)\,,
\ee
where we have used the fact that ${}_0F_1(\tfrac12,-\frac{x^2}4)=\cos x$.  
The above integral is not convergent but can be considered as a
distribution (see e.g. Chapter 6 of \cite{Estrada2012}).
By multiplying the integrand with $e^{-\s\, u}$, 
we can evaluate the integral in the region ${\rm Re}(\s)>|{\rm Re}(\a)|+|\rm{Im}(\b)|$.  
By taking the limit $\s\rightarrow 0^+$, we obtain
a finite result (hence the original integral is Abel summable \cite{Estrada2012}),
\be
	\l_{\e}(0,\b,\a)=-
	\frac{\sin\frac\a2\,
	(\cosh\frac\b2)^\frac{1+\e}2
	(\cos\frac\a2)^\frac{1-\e}2
	}{\cosh\b-\cos\a}\,.
	\label{function_lambda}
\ee
We can analytically continue the above to other region of $\a,\b$. 
Plugging this into \eqref{zeta lambda}, 
we obtain the primary contribution of the zeta function as
\be
	 \zeta_{1,[\Delta;\Y]}(z) = \int_0^\infty \frac{\dd \beta\,\b^{2z-1}}{\Gamma(2z)}\,
	 f_{1,[\Delta;\Y]}(\b),
	 \label{zeta 1 even AdS}
\ee
where the function $f_{1,[\Delta;\Y]}(\b)$ is given by
\be
	f_{1,[\Delta;\Y]}(\b)
	=\sum_{k=0}^r\,\oint_C \mu(\bm \a)\,
	   \frac{\sinh\frac\b2\,(\cosh\frac\b2)^\frac{1+\e}2
	(\cos\frac{\a_k}2)^\frac{1-\e}2}{(\cosh\b-\cos\a_k)}
	     \prod_{\substack{0 \leqslant j \leqslant r \\ j \neq
        k}}    \frac{\cosh\beta-\cos\alpha_j}{\cos\alpha_k-\cos\alpha_j}\,
 \chi^{so(2,d)}_{(\Delta ;\Y)}(\beta; \vec
    \alpha_k)\,.
    \label{f 1 even AdS}
\ee
We finally reached a formula where the relation itself does not involve
any dependency on $\Delta$ and $\Y$.

\subsubsection*{Secondary contribution}
The second term $z\,\l'_\e(0,\b,\a)$ in \eqref{lambda exp} is linear in $z$,
hence contributes to the derivative of zeta function as a residue:
\be
	 \zeta'_{2,[\Delta;\Y]}(0)=\oint \frac{\dd \beta}{2\pi i\,\b}\,
	 f_{2,[\Delta;\Y]}(\b)\,,
	 \label{res f 2}
\ee
with 
\ba
	f_{2,[\Delta;\Y]}(\b)
	\eq -\sum_{k=0}^r\oint_C \mu(\bm \a)\,
	\l'_\e(0,\b,\a_k)\times\nn
	&&\times\,
	   \frac{\sinh\frac\b2}{\sin\frac{\a_k}2}\,
	     \prod_{\substack{0 \leqslant j \leqslant r \\ j \neq
        k}}    \frac{\cosh\beta-\cos\alpha_j}{\cos\alpha_k-\cos\alpha_j}\,
 \chi^{so(2,d)}_{(\Delta ;\Y)}(\beta; \vec
    \alpha_k)\,.
    \label{f 2}
\ea
Remark that, since the second line of \eqref{f 2} is regular in $\b$,
only terms with negative powers in the Laurent series of $\l'_\e(0,\b,\a)$
can contribute to the residue in \eqref{res f 2}. 
To isolate these negative power terms, 
let us split $\tanh^{\epsilon}(\pi u)$ in the integrand of $\l_\epsilon(z,\b,\a)$
given in
 \eqref{lambda-epsilon-z}
 into two terms as
\be
\tanh^{\epsilon}(\pi u)=1-\frac 2 {1+\epsilon \,e^{2\pi u}}\,.
\label{tanh-split}
\ee
Focusing on the second term of \eqref{tanh-split}, 
the $u$ integral \eqref{lambda-epsilon-z} 
gives a regular function of $\b$, so  all the negative power terms come from the first term ``1". 
Now moving to the first term of \eqref{tanh-split},
 the corresponding integral can be evaluated for ${\rm Re}(z)>0$ as
\be
	\int_0^\infty \dd u \, \sinh(\alpha\,u)\, {}_0F_1\big(z+\tfrac12,-\tfrac{u^2\,\b^2}{4}\big)=
	(2z-1)\,\frac \a {\b^2}\,{}_2F_1\!\left(1,\tfrac 3 2-z;\tfrac 3 2; -\tfrac{\a^2}{\b^2}\right),
\ee
from which we obtain
\ba
	\l_\e'(0,\b,\a)
	\eq\frac{2\,\b\,\arctan\big(\tfrac\a\b\big)}{\b^2+\a^2}+\cO(1)\,.
	\label{lambda-prime}
\ea
Here, $\cO(1)$ means a regular function of $\b$ and it is related to the second term of \eqref{tanh-split}.
Again, both of the $u$ integrals arising from the first and second terms in \eqref{tanh-split}
are evaluated in the region where ${\rm Re}(\a)=\rm{Im}(\b)=0$,
then analytically continued to other values.
In the end, we obtain a compact 
expression of the secondary contribution:
\be
 \zeta'_{2,[\Delta;\Y]}(0)=-\oint \frac{\dd \beta}{2\pi i}\,\sum_{k=0}^r\oint_C \mu(\bm \a)\,
	\frac{2\arctan\big(\tfrac{\a_k}\b\big)}{\b^2+\a_k^2}\,
	   \frac{\sinh\frac\b2}{\sin\frac{\a_k}2}\,
	     \prod_{\substack{0 \leqslant j \leqslant r \\ j \neq
        k}}    \frac{\cosh\beta-\cos\alpha_j}{\cos\alpha_k-\cos\alpha_j}\,
 \chi^{so(2,d)}_{(\Delta ;\Y)}(\beta; \vec
    \alpha_k)\,,
    \label{zeta 2 '}
\ee
where the $\b$ contour now encloses anti-clockwise the branch cut from
$-i\,\a_k$ to $i\,\a_k$\,.  Remark that if the character is even in
$\b$, then the whole integrand is even in $\b$ and the residue vanishes trivially. 
Therefore, in such a case, the secondary contribution of the zeta function vanishes.
Furthermore, the secondary contribution is independent of $\epsilon$, i.e. whether the spectrum is bosonic or fermionic. 

\subsection{Cross-check}
\label{sec:cross_check}

In the above, we have derived the CIRZ in any dimensions.
The derivation required many technical steps, hence
it would be good if we can cross-check the formula in the end. 
In principle, if the zeta function
of a field with an arbitrary mass and spin 
is reproduced from CIRZ by inserting the character of the corresponding field,
 it will be a sufficient test for our formula.   
In this section, 
we  show how the spectral integral expressions for $\zeta'_{[\Delta,\Y]}(0)$
given in
\eqref{zeta' int} and \eqref{even AdS diff} for
odd and even dimensional AdS 
can be reproduced starting from the CIRZ formula obtained in the previous section. 
This, besides being a check, displays interesting analogies between
the spectral and Hurwitz
 zeta functions. 

\subsubsection*{AdS$_{2r+1}$}

Let us consider first the CIRZ formula   \eqref{zeta'} in odd dimensional AdS.
By inserting the $so(2,d)$ character, we obtain
\begin{eqnarray}
  \zeta'_{[\Delta;\Y]}(0) \eq \frac{\ln R}{2^{r-1}}\,\oint
  \frac{\dd\beta}{2\pi\,i}\,\sum_{k=0}^r \oint_C \mu(\bm \a)\,
  \frac{e^{-\bar\Delta\,\b}\,\chi^{so(d)}_{\Y}(\vec
    \a_k)}{(\b^2+\a_k^2)}\, \prod_{\substack{0 \leqslant j \leqslant r
      \\ j \neq k}} \frac{1}{(\cos\alpha_k-\cos\alpha_j)}\nn \eq
  -\frac{\ln R}{2^{r-1}}\,\sum_{k=0}^r \oint_C \mu(\bm \a)\,
  \frac{\sin(\a_k\,\bar \D)}{\a_k}\, \chi^{so(d)}_{\Y}(\vec \a_k)\,
  \prod_{\substack{0 \leqslant j \leqslant r \\ j \neq k}}
  \frac{1}{(\cos\alpha_k-\cos\alpha_j)}\,,
\end{eqnarray}
where $\bar\Delta=\Delta-\frac d2$ and we have performed the $\b$ integration.
From the sine function in the integrand, it is manifest that
\be
	 \zeta'_{[\frac d2;\Y]}(0)=0\,.
\ee
By taking a derivative with respect to $\Delta$, the formula becomes
\begin{equation}
  \frac{\partial}{\partial \Delta}\zeta'_{[\Delta;\Y]}(0) = -\frac{\ln
    R}{2^{r-1}}\,\sum_{k=0}^r \oint_C \mu(\bm \a)\,
  \cos(\bar\Delta\,\a_k)\,\chi^{so(d)}_{\Y}(\vec \a_k)\,
  \prod_{\substack{0 \leqslant j \leqslant r \\ j \neq k}}
  \frac{1}{(\cos\alpha_k-\cos\alpha_j)}\,,
\end{equation}
Using \eqref{def_Pd},\eqref{prop_character_so2+d} and \eqref{dim char}, one can show that
\begin{equation}
  \frac{\partial}{\partial \D} \zeta'_{[\D;\Y]}(0) = - \ln R\,
  \dim_{(-\D;\Y)}^{so(d+2)}\,.
\end{equation}
Therefore, the odd dimensional CIRZ formula \eqref{zeta'} correctly
reproduces the known result \eqref{zeta' id}.

\subsubsection*{AdS$_{2r+2}$}
\label{sec: AdS even test}

In even dimensional AdS, 
we show how the difference
of the zeta function can be obtained from the CIRZ \eqref{zeta 1 even AdS}.  In fact, the derivation shows an
interesting analogy with one well-known property of Hurwitz zeta
function:
\begin{equation}
  i^{-z}\, \zeta(z,a) + i^z\, \zeta(z,1-a) =
  \frac{(2\pi)^z}{\Gamma(z)}\, {\rm Li}_{1-z}(e^{-2\pi i\,a})\,.
  \label{zeta Li id}
\end{equation}
It can be derived from the integral representation,
\begin{equation}
  \zeta(z,a) = \int_0^\infty \frac{\dd \b\,\b^{z-1}}{\Gamma(z)} \,
  \frac{e^{-a\,\b}}{1-e^{-\b}} = -\Gamma(1-z) \oint \frac{\dd \b}{2\pi
    i} \, \frac{(-\b)^{z-1}\,
    e^{-(a-\frac12)\,\b}}{2\,\sinh\frac\b2}\,,
\end{equation}
where the integral is along the Hankel contour. For  ${\rm Re}(z)<0$ and $a\in(0,1)$, we can add to the contour an infinite
clockwise circle. Then, the contour can be deformed to encircle the
infinitely many simple poles arising from $1/\sinh\frac\b2$: they are
at $\b=2\pi\,i\,n$ for all integer $n\neq 0$. Hence, we obtain
\begin{eqnarray}
  \zeta(z,a)\eq \Gamma(1-z) \sum_{n=1}^{\infty}\left[ (-1)^n\,
    (2\pi\,i\,n)^{z-1}\, e^{(a-\frac12)\,2\pi\,i\,n} + (-1)^n\,
    (-2\pi\,i\,n)^{z-1}\, e^{-(a-\frac12)\,2\pi\,i\,n} \right] \nn \eq
  \Gamma(1-z)\, (2\pi)^{z-1} \left[ i^{z-1}\,{\rm
      Li}_{1-z}(e^{2\pi\,i\,a}) + i^{1-z}\, {\rm
      Li}_{1-z}(e^{-2\pi\,i\,a})\right],
\end{eqnarray}
from which we can easily derive \eqref{zeta Li id}. In particular, it
allows to cancel the divergence of the zeta function at $z=1$ and
evaluate the finite part:
\begin{equation}
  \lim_{z\to 1} \left[ i^{-z}\, \zeta(z,a) + i^{z}\, \zeta(z,1-a)
    \right] = 2\pi\, {\rm Li}_{0}(e^{-2\pi\,i\,a}) =- \pi\, i\,
  \frac{e^{-\pi\,i\,a}}{\sin(\pi\,a)}\,.
\end{equation}
Now let us move to the difference of the spectral zeta function.
Since the difference of the character $\chi_{(\Delta;\Y)}^{so(2,d)}(\b,\vec \a)
- \chi_{(d-\Delta;\Y)}^{so(2,d)}(\b,\vec\a)$ is even in $\b$,
the corresponding secondary contribution simply vanishes:
$\zeta'_{2,[\Delta;\Y]}(0)-\zeta'_{2,[d-\Delta;\Y]}(0)=0$.
Focussing on the primary contribution \eqref{zeta 1 even AdS},  we first find that
the integrand function $f_{1,[\Delta;\Y]}$ reduces to
\begin{equation}
  f_{1,[\Delta;\Y]}(\b) = \sum_{k=0}^r \oint_C \mu(\bm\a)\,
  \frac{\chi^{so(d)}_\Y(\vec\a_k)}{2^{r+1}\, \Pi_{\substack{0
        \leqslant j \leqslant r \\ j \neq k}} (\cos\a_k-\cos\a_j)}\,
  \frac{(\cosh\frac\b2)^\frac{1+\e}2 (\cos\frac{\a_k}2)^\frac{1-\e}2\,
    e^{-\bar\Delta\,\b}}{\cosh\b-\cos\a_k}\,,
\end{equation}
by inserting the expression of $\chi^{so(2,d)}_{(\Delta;\Y)}$.
By changing the order of the $\b$ and $\bm\a$ integrals:  
\be
	\zeta_{1,[\Delta,\Y]}(z)=
	\sum_{k=0}^r\oint_C \mu(\bm\a)\,\frac{\chi^{so(d)}_\Y(\vec\a_k)\,\xi(2z,\bar\Delta,\a_k)}{2^{r+1}\,
	\Pi_{\substack{0 \leqslant j \leqslant r \\ j \neq
        k}}    (\cos\a_k-\cos\a_j)}\,,
        \label{zeta e t}
\ee
with
\be
	\xi(z,\bar\Delta,\a)=\int_0^\infty \frac{\dd\b\,\b^{z-1}}{\Gamma(z)}\,
	\frac{(\cosh\frac\b2)^\frac{1+\e}2
	(\cos\frac{\a}2)^\frac{1-\e}2\,e^{-\bar\Delta\,\b}}{\cosh\b-\cos\a}\,,
	\label{xi int}
\ee
we can focus first on the function $\xi(z,\bar\Delta)$ which plays
an analogous role as the Hurwitz zeta function in the relation \eqref{zeta Li id}.
In particular, $\frac{\partial}{\partial\D}\xi(z,\bar\D,\a)=-z\,\xi(z+1,\bar\D,\a)$.
We recast the  integral \eqref{xi int} as the integral over the Hankel contour,
\be
	\xi(z,\bar\Delta,\a)=-\Gamma(1-z)\,\oint \frac{\dd\b\,(-\b)^{z-1}}{2\pi i}\,
	\frac{(\cosh\frac\b2)^\frac{1+\e}2
	(\cos\frac{\a}2)^\frac{1-\e}2\,e^{-\bar\Delta\,\b}}{2\,\sinh\frac{\b+i\,\a}2\,\sinh\frac{\b-i\,\a}2}\,.
\ee
Like in the case of Hurwitz zeta function, we add to the contour an infinite clockwise circle
and shrink it to enclose the infinite many simple poles arising, at this time, from
$\sinh\frac{\b\pm i\,\a}2$:
they are at $\b=(2\pi\,n\mp \a)\,i$\,.  By collecting the residues, we get
\be
	\xi(z,\bar\Delta,\a)=
	\frac{\Gamma(1-z)\,e^{\bar\Delta\,\a\,i}}{2\,i\,\sin\frac\a2}\,
	\sum_{n=-\infty}^\infty 
	[-(2\pi\,n-\a)\,i]^{z-1}\,(-\e)^n\,e^{-2\pi\,n\,\bar\Delta\,i}+(\a\leftrightarrow -\a)\,,
\ee
which is divergent in the $z\to 1$ limit.
The divergence can be canceled by taking the difference,
\ba
	\lim_{z\to1}\left[\xi(z,\bar\Delta,\a)-\xi(z,-\bar\Delta,\a)\right]
	\eq\frac{\pi}2\,\frac{e^{\bar\Delta\,\a\,i}}{\sin\frac\a2}\,\tanh^\e(\pi\,i\,\bar\Delta)
	+(\a\leftrightarrow -\a)\nn
	\eq \pi\,i\,\frac{\sin(\bar\Delta\,\alpha)}{\sin\frac{\a}2}\,\tanh^\e(\pi\,i\,\bar\Delta)\,.
\ea
Using the above result in \eqref{zeta e t} 
together with \eqref{def_Pd}, \eqref{prop_character_so2+d} and \eqref{dim char}, 
we can reproduce	\eqref{even AdS diff}.

\section{Derivative expression of the CIRZ}
\label{sec:derivative_expansion}

In this section, we present a different expression of CIRZ in terms of
derivatives in $\a_i$. As explained in
\hyperref[sec:zeta_function_AdS]{Section \ref{sec:zeta_function_AdS}},
the combination $\dim_\Y^{so(d)} \mu_\Y(u)$ in the zeta function
\eqref{zeta_camporesi_higuchi} is related to the Weyl dimension
formula which can be obtained as a limit of the $so(d+2)$ characters.
In \hyperref[sec:CIRZ]{Section \ref{sec:CIRZ}}, this limit was taken
as a contour integral.  The expression with contour integrals in
$\a_i$ variables is useful --- see the companion paper \cite{part_ii}
--- but sometimes not explicit enough.  For instance, if one wants to
implement the CIRZ formula in a computer program, it will be more
convenient to have an expression, where all the $\a_i$ contour
integrals are already evaluated using the residue theorem, involving
$\a_i$ derivatives of the $so(d+2)$ characters. In fact, for an
expression in terms of $\a_i$ derivatives, it is simpler to re-derive
the CIRZ by taking the limit of $so(d+2)$ characters using a
generalized L'H\^opital's rule. Below, we demonstrate how to obtain
such an expression. The CIRZ for AdS$_4$ and AdS$_5$ originally
presented in \cite{Bae:2016rgm} are recovered as special cases.

\subsection{General dimensions}

In order to recover the Weyl character formula
\eqref{Weyl_dim_formula}, we need to evaluate the $so(d+2)$ characters
in the limit $\bm \alpha \to \bm 0$.  It is actually subtle to perform
the evaluation since
the $so(d+2)$ characters take
the form,
\begin{equation}\label{eq:charfrac}
  \chi^{so(d+2)}_{(s_0,\Y)}(\bm \alpha) =
  \frac{\mathsf{N}_{(s_0,\Y)}^{(d+2)}(\bm
    \alpha)}{\mathsf{D}^{(d+2)}(\bm \alpha)}\, ,
\end{equation}
and  both the
numerator and denominator vanish as $\bm\alpha\to\bm 0$\,:
\begin{equation}
  \lim_{\bm \alpha \rightarrow \bm 0}
  \mathsf{N}_{(s_0,\Y)}^{(d+2)}(\bm \alpha) = \lim_{\bm \alpha
    \rightarrow \bm 0} \mathsf{D}^{(d+2)}(\bm \alpha) = 0\, .
\end{equation}
The explicit expressions for the numerator
$\mathsf{N}_{(s_0,\Y)}^{(d+2)}$ and the denominator
$\mathsf{D}^{(d+2)}$ are given in \eqref{num_char_1} and
\eqref{denominator}. 
Despite the apparent singularity, the limit of the $so(d+2)$ character
does exist, and can be obtained by using a generalisation of
L'H\^opital's rule (see \hyperref[sec: app B]{Appendix B} for more
details):
\begin{equation}\label{eq:charlimit}
  \dim_{(s_0,\Y)}^{so(d+2)} = \lim_{\bm \alpha \rightarrow \bm 0}
  \chi^{so(d+2)}_{(s_0,\Y)}(\bm \alpha) =
  \frac{\mathscr{D}_{\Phi^{d+2}}\, \mathsf{N}_{(s_0,\Y)}^{(d+2)}(\bm
    \alpha) \big|_{\bm \alpha = \bm 0}}{\mathscr{D}_{\Phi^{d+2}}\,
    \mathsf{D}^{(d+2)}(\bm \alpha) \big|_{\bm \alpha = \bm 0}}\, ,
\end{equation}
where the differential operator $\mathscr D_{\Phi^{d+2}}$ is given, in
the notation $\partial_i=\partial_{\alpha_i}$, by
\begin{equation}\label{D def}
  \mathscr D_{\Phi^{d+2}} =  \prod_{0 \leqslant i < j \leqslant r}
  (\partial_i^2-\partial_j^2)\, \times \, \left\{
  \begin{aligned}
    1 \quad & \qquad [d=2r]\\ \, \, \prod_{k=0}^r \partial_k
    & \qquad [d=2r+1]
  \end{aligned}
  \right.\,.
\end{equation}
Firstly, the denominator of \eqref{eq:charlimit} depends only on
$d$\,:
\begin{equation}
  \mathscr{D}_{\Phi^{d+2}}\, \mathsf D^{(d+2)}(\bm \alpha) \Big|_{\bm
    \alpha = \bm 0}=c_d
  \label{c d}
\end{equation}
and can be explicitly evaluated as explained in \hyperref[sec:
  den]{Appendix \ref{sec: den}}. The result reads
\begin{equation}
  c_d = (-1)^{{}^{{}_{\frac {r(r+1)} 2}}} 2^r (r+1)!
  \prod_{0\leqslant i < j \leqslant r} (d-i-j)(j-i) \times \left\{
  \begin{aligned}
     1 \qquad \qquad & \qquad [d=2r]\\ 2\, i^{r+1}\, \prod_{k=0}^r
     (d/2-k) & \qquad [d=2r+1]
  \end{aligned}
  \right..
  \label{formula_c}
\end{equation}
Secondly, the numerator of \eqref{eq:charlimit} can be recast, as
explained in \hyperref[sec: num]{Appendix \ref{sec: num}}, into
\begin{equation} \label{DN}
  \mathscr{D}_{\Phi^{d+2}}\, \mathsf{N}_{(s_0,\Y)}^{(d+2)}(\bm \alpha)
  \big|_{\bm \alpha = \bm 0}= 2(r+1) \,(-1)^{d}\, u^{d-2r}\,
  \sum_{n=0}^r u^{2n}\, \mathfrak{D}_{(n)} \mathsf N_\Y^{(d)}(\vec
  \alpha)\Big|_{\bm\alpha=\bm 0},
\end{equation}
where the differential operator $\mathfrak{D}_{(n)}$ is defined as
\begin{equation}
  \mathfrak{D}_{(n)} := \bar \partial_{(n)} \mathscr{D}_{\Phi^{d}}\,,
  \label{differential_operator}
\end{equation}
with 
\begin{equation}
  \bar \partial_{(n)} = (-1)^{r-n}\, \sum_{1\leq i_1<i_2<\dots<
    i_{r-n}\leq r}\partial_{i_1}^2 \dots \partial_{i_{r-n}}^2\,.
\end{equation}
Using \eqref{c d} and \eqref{DN}, we can write the zeta function as
\begin{equation}\label{eq:zetachar}
  \zeta_{[\Delta;\Y]}(z) =\frac {2(r+1)}{c_d} \,(-1)^{d}\,
  \sum_{n=0}^r\, \int_0^\infty \frac{\dd u\,\rho_{\e}(u)}{\big[u^2 +
      (\Delta-\frac d2)^2 \big]^z}\, u^{2n+d-2r}\,
  \mathfrak{D}_{(n)}\, \mathsf N_\Y^{(d)}(\vec \alpha)\Big|_{\vec
    \alpha=\vec 0} \,,
\end{equation}
where $\rho_\epsilon(u)$ is given in \eqref{def_rho}.  At this point,
introducing the $\beta$-integral \eqref{prop_integral_beta} will give
rise to a factor $e^{-\beta\,\Delta}$.  By reconstructing the
$so(d+2)$ character from $e^{-\beta\,\Delta}$ and $\mathsf N_\Y^{(d)}$
and using the identity \eqref{DD}, we obtain
\begin{eqnarray}
  \zeta_{[\Delta;\Y]}(z) & = & \frac{2(r+1)}{\Gamma(z)\, c_d}
  \int_0^\infty \dd \beta\, \sum_{n=0}^r\, \varphi_{\e,n}(z,\beta)\,
  \mathfrak{D}_{(n)} \Big[ \mathsf{D}^{(d+2)}(i\beta, \vec \alpha)\,
    \chi^{so(2,d)}_{(\Delta,\Y)}(\beta,\vec \alpha) \Big]\Big|_{\vec
    \alpha=\vec 0}\,,\quad
  \label{zeta_diff_fin}
\end{eqnarray}
with
\begin{equation}
  \varphi_{\e,n}(z,\beta) = \sqrt{\pi}\, \int_0^\infty \dd u\,
  \left( \frac{\beta}{2u}
  \right)^{z-\frac12}\, J_{z-\frac12}(\beta\,u)\, \times \left\{
  \begin{aligned}
    \frac{\ln R}{\pi}\,u^{2n} \qquad & \quad [d=2r]\\ 
    \frac{i}2\,\tanh^\e(\pi\,u)\,u^{2n+1}\, &\quad
    [d=2r+1]
  \end{aligned}
  \right.\,.
\end{equation}
Remark that the function $\varphi_{\e,n}(z,\beta)$ is related
to $\nu_\e(z,\b,\a)$ defined in \eqref{nu} as
\begin{equation}
  \varphi_{\e,n}(z,\beta) = \frac{\sqrt{\pi}}2\times\left\{
  \begin{aligned}
    \partial_\a^{2n}\,\nu_\e(z,\b,\a)\,\big|_{\a=0} & \qquad [d=2r]
    \\ -i\,\partial_\a^{2n+1}\,\nu_\e(z,\b,\a)\,\big|_{\a=0}&\qquad
       [d=2r+1]
  \end{aligned}
  \right.\,.
\end{equation}
Hence, the expression \eqref{zeta_diff_fin} can be considered as the
result of the $\alpha_i$ contour integrals of \eqref{zeta intmed}.

\subsubsection*{Odd dimensional AdS}
For $d=2r$, the function $\nu_\e(z,\b,\a)$ has been computed exactly
as \eqref{nu even d}. The corresponding $\varphi_{\e,n}(\beta;z)$ is
\begin{equation}
  \varphi_{\e,n}(z,\beta) =\frac{\ln R}{2\,\sqrt{\pi}}\,
  \frac{\Gamma(n+\tfrac12)}{\Gamma(z-n)}\,
  \Big(\frac\beta2\Big)^{2(z-n-1)}\,,
\end{equation}
hence, the zeta function can be expressed as
\begin{eqnarray}
  \zeta_{[\Delta;\Y]}(z) & = & \frac{\ln R}{\Gamma(z)}\,\frac{d+2}{2\,
    c_d} \int_0^\infty \dd \beta\, \sum_{n=0}^r\, \frac{(2n)!}{4^n
    n!}\, \frac{\big(\tfrac\beta2\big)^{2(z-n-1)}}{\Gamma(z-n)}\,
  \times \nn && \qquad \qquad \qquad \qquad \qquad  \times\,
  \mathfrak{D}_{(n)} \Big[ \mathsf{D}^{(d+2)}(i\beta, \vec \alpha)\,
    \chi^{so(2,d)}_{(\Delta,\Y)}(\beta,\vec \alpha) \Big]\Big|_{\vec
    \alpha=\vec 0}\,.\quad
\end{eqnarray}
The $\b$ integral is convergent for ${\rm Re}(z)>d$, but one can
analytically continue $z$ to other values. Since the only singularity
of the integrand is at $\beta=0$, we can deform the $\b$ integral to a
complex integral with the contour encircling the origin
counter-clockwise:
\begin{equation}
  \int_0^\infty \dd\beta\,
  \frac{\big(\tfrac\beta2\big)^{2(z-1-n)}}{\Gamma(z-n)}\, f(\beta) =
  (-1)^n\, 2^{2n+1}\, n!\, \oint \frac{\dd\beta}{2\,\pi\,i}\,
  \frac{f(\beta)}{\beta^{2(n+1)}} + {\cal O}(z)\,.
\end{equation}
In the end, the first derivative of the zeta function in AdS$_{2r+1}$
reads
\begin{equation}
  \frac{\zeta'_{[\Delta,\Y]}(0)}{\ln\,R} = \frac{d+2}{c_d} \oint
  \frac{\dd\beta}{2\,\pi\,i}\, \sum_{n=0}^r\,
  \frac{(-1)^n\,(2n)!}{\beta^{2(n+1)}}\, \mathfrak{D}_{(n)} \Big[
    \mathsf{D}^{(d+2)}(i\beta, \vec \alpha)\,
    \chi^{so(2,d)}_{(\Delta,\Y)}(\beta,\vec \alpha) \Big]\Big|_{\vec
    \alpha=\vec 0}\,,
\end{equation}
which contains the $\a_i$-derivatives instead of the contour integrals
in \eqref{f even}.

\subsubsection*{Even dimensional AdS}
For $d=2r+1$,  the functions $\varphi_{\e,n}(z,\b)$ is given by
\begin{equation}
  \varphi_{\e,n}(z,\b) = \frac{i\, \sqrt{\pi}\,\big(\tfrac
    \beta2\big)^{2z-1}}{2\,\Gamma(z+\tfrac12)}\,
  \mu_{\e,n}(z,\beta)\, ,
\end{equation}
with
\begin{equation}
  \mu_{\e,n}(z,\beta) = \partial_\alpha^{2n+1}
  \lambda_\epsilon(z,\beta,\alpha)\,\big|_{\alpha=0}\,.
\end{equation}
Following \hyperref[sec: odd d CIRZ]{Section \ref{sec: odd d CIRZ}},
we focus on the first two Taylor coefficients of
$\lambda_\epsilon(z,\beta,\alpha)$ in $z$, which have been computed in
\eqref{function_lambda} and \eqref{lambda-prime}. They immediately
give $\mu_{\e,n}(0,\b)$ and
$\mu'_{\e,n}(0,\b)=\frac{\partial}{\partial
  z}\,\mu_{\e,n}(z,\b)\,|_{z=0}$ corresponding to the primary and
secondary contributions to the zeta function.  The
primary contribution reads
\begin{equation}
  \zeta_{[\Delta,\Y],1}(z) = \frac{i\, (d+1)}{2\,c_d} \int_0^\infty
  \dd \beta\, \frac{\beta^{2z-1}}{\Gamma (2z)\,}\, \sum_{n=0}^r\,
  \mu_{\e,n}(0,\beta)\, \mathfrak{D}_{(n)} \Big[
    \mathsf{D}^{(d+2)}(i\beta, \vec \alpha)\,
    \chi^{so(2,d)}_{(\Delta,\Y)}(\beta,\vec \alpha) \Big]\Big|_{\vec
    \alpha=\vec 0}\,.
\end{equation}
and the secondary contribution is 
\begin{equation}
  \zeta'_{[\Delta,\Y],2}(0)= \frac{i\, (d+1)}{2\,c_d} \oint \frac{\dd
    \beta}{2\pi\, i\, \beta}\, \sum_{n=0}^r\, \mu'_{\e,n}(0,\beta) \,
  \mathfrak{D}_{(n)} \Big[ \mathsf{D}^{(d+2)}(i\beta, \vec \alpha)\,
    \chi^{so(2,d)}_{(\Delta,\Y)}(\beta,\vec \alpha) \Big]\Big|_{\vec
    \alpha=\vec 0}\,.
\end{equation}

\subsection{Explicit expressions in low dimensions}
In this section we spell out the explicit formulae for the zeta
function in AdS$_3$, AdS$_5$ and AdS$_7$, and its primary and
secondary contributions in AdS$_4$ and AdS$_6$. In order to display
the various formulae in a compact way, let us introduce
\begin{equation}
  \mathsf{f}_{(\Delta,\Y)}^{d,(n)}(\beta) := \mathfrak{D}_{(n)} \Big[
    \mathsf{D}^{(d+2)}(i\beta, \vec \alpha)\,
    \chi^{so(2,d)}_{(\Delta,\Y)}(\beta,\alpha) \Big]\Big|_{\alpha=0} \,
  \times \left\{
  \begin{aligned}
    \frac{(d+2)\, (2n)!}{2^{2n+1}\, n!\, c_d} & \quad
         [d=2r]\\ \frac{i\,(d+1)}{2\, c_d} \quad & \quad [d=2r+1]
  \end{aligned}
  \right.\,.
  \label{def_f}
\end{equation}
Then, the zeta function in AdS$_{2r+1}$ reads
\begin{eqnarray}
  \zeta_{[\Delta;\Y]}(z) & = & \frac{\ln R}{\Gamma(z)}\, \int_0^\infty
  \dd \beta\, \sum_{n=0}^r\,
  \frac{\big(\tfrac\beta2\big)^{2(z-n-1)}}{\Gamma(z-n)}\,
  \mathsf{f}_{(\Delta,\Y)}^{2r,(n)}(\beta)\,,
  \label{zeta_f_even}
\end{eqnarray}
whose first derivative is given by
\begin{equation}
  \zeta'_{[\Delta,\Y]}(0) = \ln\,R\, \oint \frac{\dd\beta}{2\,\pi\,i}\,
  \sum_{n=0}^r\, (-1)^n\, \frac{2^{2n+1}\,n!}{\beta^{2(n+1)}}\,
  \mathsf{f}_{(\Delta,\Y)}^{2r,(n)}(\b)\,.
\end{equation}
On the other hand, the primary contribution to the zeta function in
AdS$_{2r+2}$ reads
\begin{eqnarray}
  \zeta_{[\Delta;\Y],1}(z) & = & \int_0^\infty \dd \beta\,
  \frac{\beta^{2z-1}}{\Gamma(2z)}\, \sum_{n=0}^r\,
  \mu_{\epsilon,n}(0,\beta)\,
  \mathsf{f}_{(\Delta,\Y)}^{2r+1,(n)}(\beta)\,,
  \label{zeta_f_odd}
\end{eqnarray}
whereas the secondary contribution is given by
\begin{equation}
  \zeta'_{[\Delta;\Y],2}(0) = \oint \frac{\dd \beta}{2\pi\, i\,
    \beta}\, \sum_{n=0}^r\,
  \mu'_{\e,n}(0,\beta) \,
  \mathsf{f}_{(\Delta,\Y)}^{2r+1,(n)}(\beta)\,.
  \label{secondary_derivative}
\end{equation}
To derive the explicit expressions below, we will use
\eqref{formula_c}, \eqref{differential_operator} and
\eqref{denominator}.

\subsubsection{AdS$_3$}
In order to write down explicitly the functions $\mathsf{f}_{\cal
  H}^{2,(0)}(\beta)$ and $\mathsf{f}_{\cal H}^{2,(1)}(\beta)$ relevant
to the computation of the zeta function in AdS$_3$, we need to know
the expression of the differential operator 
\begin{equation}
  \mathfrak{D}_{(0)}= \bar \partial_{(0)}\,\mathscr{D}_{\Phi^2}
  ,\qquad \mathfrak{D}_{(1)} = \bar
  \partial_{(1)}\,\mathscr{D}_{\Phi^2}\,\,.
\end{equation}
Since $\mathscr{D}_{\Phi^2}$ is the identity\,\footnote{Indeed,
  $so(2)$ being unidimensional, it does not have a root space
  decomposition like the higher dimensional orthogonal algebras.}, we
have
\begin{equation}
  \mathfrak{D}_{(0)} = -\partial_\alpha^2\, , \qquad
  \mathfrak{D}_{(1)} = 1\, .
\end{equation}
Together with
\begin{equation}
  c_2 = -4\, , \qquad \mathsf{D}^{(4)}(i\beta,\alpha) = 2\,
  (\cosh\beta-\cos\alpha)\,,
\end{equation}
we find
\begin{eqnarray}
  \mathsf{f}_{\cal H}^{2,(0)}(\beta) & = & \Big( 1 + 2
  \sinh^2\tfrac\beta2\, \partial_\alpha^2\, \Big)\,
  \chi^{so(2,2)}_{\cal H}(\beta,\alpha)\big|_{\alpha=0}\,,
\end{eqnarray}
and
\begin{equation}
  \mathsf{f}_{\cal H}^{2,(1)}(\beta) = - \sinh^2\tfrac\beta2
  \ \chi^{so(2,2)}_{\cal H}(\beta,0)\,.
\end{equation}
Inserting these ingredients into \eqref{zeta_f_even}, we obtain
\begin{equation}
  \zeta_{\cal H}(z) = \frac{\ln R}{\Gamma(z)^2}\, \int_0^\infty \dd
  \beta\, \Big(\frac\beta2\Big)^{2(z-1)}\, \Big[ 1\, +
    \tfrac{4(1-z)}{\beta^2}\,\sinh^2\tfrac\beta2 + 2
    \sinh^2\tfrac\beta2\, \partial_\alpha^2 \Big] \chi^{so(2,2)}_{\cal
    H}(\beta,\alpha)\big|_{\alpha=0}\,,
\end{equation}
and
\begin{equation}
  \zeta'_{\cal H}(0) = \ln \,R\, \oint \frac{\dd\beta}{2\,\pi\,i}\,
   \frac 2 {\beta^2}\, \Big(  1\, +
    \tfrac{4}{\beta^2}\,\sinh^2\tfrac\beta2 + 2
    \sinh^2\tfrac\beta2\, \partial_\alpha^2 \Big)\,
  \chi^{so(2,2)}_{\cal H}(\beta,\alpha)\big|_{\alpha=0}\,.
\end{equation}

\subsubsection{AdS$_4$}
The relevant differential operators read
\begin{equation}
  \mathscr{D}_{\Phi^3} = \partial_\alpha\, , \qquad \bar
  \partial_{(0)} = -\partial_\alpha^2\, , \qquad \bar \partial_{(1)} =
  1\, ,
\end{equation}
Using
\begin{equation}
  c_3 = 12\,, \qquad  \mathsf{D}^{(5)}(i\beta,\alpha) =
  -8\,i\,\sinh\tfrac\beta2\,\sin\tfrac{\alpha}2\, \big( \cosh\beta -
  \cos\alpha \big)\,.
\end{equation}
we obtain
\begin{equation}
  \mathsf{f}_{\cal H}^{3,(0)}(\beta) = \tfrac13\, \sinh\tfrac\beta2
  \Big( \sinh^2\tfrac\beta2-6 - 12\sinh^2\tfrac\beta2\, \partial_\alpha^2
  \Big)\, \chi^{so(2,3)}_{\cal H}(\beta,\alpha)\big|_{\alpha=0}\,,
\end{equation}
and
\begin{equation}
  \mathsf{f}_{\cal H}^{3,(1)}(\beta) = \frac43 \sinh^3\tfrac\beta2\,
  \chi^{so(2,3)}_{\cal H}(\beta,0)\,.
\end{equation}
The functions $\mu_{\epsilon,0}$ and $\mu_{\epsilon,1}$ read
\begin{equation}
  \mu_{\e,0}(0,\beta) = -
  \frac{(\cosh\tfrac\beta2)^{\tfrac{1+\epsilon}2}}{4\,
    \sinh^2\tfrac\beta2} \, , \qquad \mu_{\e,1}(0,\beta) =
  \frac{(\cosh\tfrac\beta2)^{\tfrac{1+\epsilon}2}
    \big((5-3\,\epsilon)\sinh^2\tfrac\beta2+12\big)} {32
    \sinh^4\tfrac\beta2}\,.
  \label{mu_b}
\end{equation}
According to \eqref{zeta_f_odd}, the primary contribution is 
\begin{equation}
  \zeta_{{\cal H},1}(z) = \int_0^\infty \dd\beta\,
  \frac{\beta^{2z-1}}{\Gamma(2z)}\,
  \frac{(\cosh\tfrac\beta2)^{\tfrac{1+\epsilon}2}}{\sinh \tfrac \b
    2}\, \left( 1+\tfrac{1-\epsilon} 8\,\sinh^2\tfrac \b 2+\sinh^2
  \tfrac \b 2\partial_\a^2 \right)\, \chi^{so(2,3)}_{\cal
    H}(\beta,\alpha)\big|_{\alpha=0}\,,
\end{equation}
By setting $\epsilon=+1$ or $-1$, we obtain the formulae derived in
\cite{Bae:2016rgm} and \cite{Pang:2016ofv} for bosonic and fermionic
spectrum, respectively. To compute the secondary contribution, we need
the first derivative of $\mu_{\e,n}(z,\beta)$:
\begin{equation}
  \mu'_{\epsilon,0}(0,\beta) = \frac{2}{\beta^2} + {\cal O}(1)\,,
  \qquad \mu'_{\epsilon,1}(0,\beta) = -\frac{16}{\beta^4} + {\cal
    O}(1)\,. \label{mu-epsilon-prime}
\end{equation}
Inserting the above ingredients into \eqref{secondary_derivative}, we
arrive at
\begin{equation}
  \zeta'_{{\cal H},2}(0) = \oint \frac{\dd \beta}{2\pi\, i}\,
  \frac{2\sinh^3\tfrac\beta2}{\beta^3}\, \left( -\frac{32}{3\b^2}-
  \frac 2{\sinh^2\tfrac\beta2} + \frac{1}{3} - 4\, \partial_\alpha^2
  \right)\, \chi^{so(2,3)}_{\cal H}(\beta,\alpha)\Big|_{\alpha=0}\,.
\end{equation}
Notice that, as we already pointed out, since the function of $\beta$
multiplying the character in the integrand is even, this secondary
contribution vanishes if $\chi^{so(2,3)}_{\cal H}(\beta,\alpha)$ is
also an even function of $\beta$.

\subsubsection{AdS$_5$}
The differential operators $\mathfrak{D}_{(n)}$ are composed of
\begin{equation}
  \mathscr{D}_{\Phi^4} = \partial_{1}^2 -
  \partial_{2}^2\, , \qquad \bar \partial_{(0)} =
  \partial_{1}^2\, \partial_{2}^2\, , \qquad \bar
  \partial_{(1)} = -\big( \partial_{1}^2 +
  \partial_{2}^2 \big),\qquad \bar \partial_{(2)} = 1\, .
\end{equation}
After some computations, one obtains
\begin{eqnarray}
  \mathsf{f}_{\cal H}^{4,(0)}(\beta) & = & \Big[ 1 - 
    \sinh^2\tfrac\beta2\, (\tfrac13\sinh^2 \tfrac \b 2-1)\,(\partial_{1}^2 +
    \partial_{2}^2) \\ && \quad - \tfrac13\, \sinh^4\tfrac\beta2\,
    \big(\partial_{1}^4+
    \partial_{2}^4 -12\, \partial_{1}^2\,\partial_{2}^2 \big) \Big]\, \chi^{so(2,4)}_{\cal
    H}(\beta,\vec\alpha)\big|_{\vec\alpha=\vec0}\,\,, \nonumber
\end{eqnarray}
\begin{equation}
  \mathsf{f}_{\cal H}^{4,(1)}(\beta) = 
  \sinh^2\tfrac\beta2\, \Big[ \tfrac 1 3 \sinh^2 \tfrac \b 2-1 - 
    \sinh^2\tfrac\beta2\,(\partial_{1}^2 + \partial_{2}^2) \Big]\,
  \chi^{so(2,4)}_{\cal
    H}(\beta,\vec\alpha)\big|_{\vec\alpha=\vec0}\,\,,
\end{equation}
and
\begin{equation}
  \mathsf{f}_{\cal H}^{4,(2)}(\beta) = \tfrac12\,
  \sinh^4\tfrac\beta2\, \chi^{so(2,4)}_{\cal H}(\beta, \vec0)\,.
\end{equation}
Plugging these expressions into \eqref{zeta_f_even}, we reproduce the CIRZ
formula for AdS$_5$ derived in \cite{Bae:2016rgm}.

\subsubsection{AdS$_6$}
To define the differential operators $\mathfrak{D}_{(n)}$, 
we need the following building blocks:
\begin{equation}
  \mathscr{D}_{\Phi^5} = \big(\partial_{1}^2 -
  \partial_{2}^2\big)\,
  \partial_{1}\partial_{2}\, , \quad \bar \partial_{(0)}
  = \partial_{1}^2\, \partial_{2}^2\, , \quad \bar
  \partial_{(1)} = -\big( \partial_{1}^2 +
  \partial_{2}^2 \big)\, , \quad \bar \partial_{(2)} = 1\, .
\end{equation}
Then we find
\begin{eqnarray}
  \mathsf{f}_{\cal H}^{5,(0)}(\beta) & = & \sinh\tfrac\beta2\, \Big[
    -2+\tfrac 1 {3}\sinh^2\tfrac\beta2 -\tfrac 3
    {20}\sinh^4\tfrac\beta2+ \tfrac 1 2\sinh^2\tfrac\beta2\,
    (\cosh\b-5)\,(\partial_{1}^2+\partial_{2}^2) \nn && \qquad \qquad
    \qquad +\tfrac 2 {3}\sinh^4\tfrac\beta2\, (\partial_{1}^4 +
    \partial_{2}^4-12\,\partial_{1}^2\,\partial_{2}^2) \Big]\,
  \chi^{so(2,5)}_{\cal
    H}(\beta,\vec\alpha)\big|_{\vec\alpha=\vec0}\,\,, \quad
\end{eqnarray}
\begin{eqnarray}
  \mathsf{f}_{\cal H}^{5,(1)}(\beta) & = & \tfrac 1 3
  \sinh^3\tfrac\beta2\, \Big(5-\cosh \b
  +4\sinh^2\tfrac\beta2 (\partial_{1}^2 +
  \partial_{2}^2) \Big)\, \chi^{so(2,5)}_{\cal
    H}(\beta,\vec\alpha)\big|_{\vec \alpha=\vec0}\,\,,\quad
\end{eqnarray}
\begin{equation}
  \mathsf{f}_{\cal H}^{5,(2)}(\beta) = -\tfrac{4}{15}\,
  \sinh^5\tfrac\beta2\, \chi^{so(2,5)}_{\cal H}(\beta, \vec0)\,.
\end{equation}
The functions $\mu_{\e,0}(0,\beta)$ and $\mu_{\e,1}(0,\beta)$ were
already computed in \eqref{mu_b}, whereas
\begin{equation}
  \mu_{\e,2}(0,\beta) = \frac{(\cosh\tfrac\b
    2)^{\tfrac{1+\epsilon}2}}{256\,\sinh^6\tfrac\beta2}
  \Big[480+120(3-\e)
    \sinh^2\tfrac\beta2+(19-30\,\e+15\,\e^2)\sinh^4\tfrac\beta2\Big]\,.
\end{equation}
The primary contribution is 
\begin{eqnarray}
  \zeta_{{\cal H},1}(z) &=& \int_0^\infty \dd\beta\,
  \frac{\beta^{2z-1}}{\Gamma(2z)}\, \frac{(\cosh\tfrac\b
    2)^{\tfrac{1+\epsilon}2}}{64\sinh\tfrac\beta2}\,
  \Big[96+(1-\e)\Big(16-(3+\e)\sinh^2\tfrac \b 2\Big)\,\sinh^2\tfrac
    \b 2\nn &&\qquad\qquad +\tfrac 8 3\Big (24- (1 + 3\, \e)
    \sinh^2\tfrac \b 2\Big)\,\sinh^2\tfrac \b 2\, (\partial_{1}^2
    +\partial_{2}^2) \nn&&\qquad\qquad -\tfrac {32}3 \,\sinh^4\tfrac
    \b 2\,(\partial_{1}^4+ \partial_{2}^4-12\,
    \partial_{1}^2\,\partial_{2}^2 ) \Big]\, \chi^{so(2,5)}_{\cal
    H}(\beta,\vec\alpha)\big|_{\vec\alpha=\vec0}\,\,.
\end{eqnarray}
Using \eqref{mu-epsilon-prime} and
\begin{equation}
  \m'_{\e,2}(0,\beta) = \frac{368}{\beta^6}\, + {\cal O}(1)\,,
\end{equation}
we obtain the secondary contribution as
\begin{eqnarray}
  \zeta'_{{\cal H},2}(0) & = & \oint \frac{\dd \beta}{2\pi\, i}\,
  \tfrac{\sinh\tfrac\beta2}{\beta^3}\, \Big( -4+\tfrac 2 3
  \sinh^2\tfrac\beta2 -\tfrac 3 {10}\sinh^4\tfrac\beta2 +
  \sinh^2\tfrac\beta2\,
  (\cosh\b-5)(\partial_{\alpha_1}^2+\partial_{\alpha_2}^2) \nn &&
  \qquad \qquad\qquad\,\,\, +\tfrac 4 3\sinh^4\tfrac\beta2\,
  (\partial_{\alpha_1}^4+
  \partial_{\alpha_2}^4-12\,\partial_{\alpha_1}^2\partial_{\alpha_2}^2
  ) \nn && \qquad \qquad\qquad\,\,\, + \tfrac{16\,
    \sinh^2\tfrac\beta2}{3\,\beta^2}\, \big( \cosh\b-5
  -4\sinh^2\tfrac\beta2 (\partial_{\alpha_1}^2 +
  \partial_{\alpha_2}^2) \big)\, \nonumber \\ && \qquad \qquad
  \qquad\,\,\, - \tfrac{1472 \,\sinh^4\tfrac\beta2}{15\,\beta^4}
  \Big)\, \chi^{so(2,5)}_{\cal
    H}(\beta,\vec\alpha)\big|_{\vec\alpha=\vec0}\,.
\end{eqnarray}

\subsubsection{AdS$_7$}
Finally, the relevant differential operators for
AdS$_7$ zeta functions is obtained by combining the differential
operator associated to the (positive) root system of $so(6)$,
\begin{eqnarray}
  \mathscr{D}_{\Phi^6} & = & \big(\partial_{1}^2 -
  \partial_{2}^2\big)\, \big(\partial_{1}^2 - \partial_{3}^2 \big)\,
  \big(\partial_{2}^2 - \partial_{3}^2\big) \,,
\end{eqnarray}
with
\begin{equation}
  \bar \partial_{(0)} = -\partial_{1}^2\, \partial_{2}^2\,
  \partial_{3}^2\, ,\quad \bar \partial_{(1)} =
  \partial_{\alpha_1}^2\partial_{2}^2 + \partial_{1}^2\partial_{3}^2 +
  \partial_{2}^2\partial_{3}^2\,,\quad \bar \partial_{(2)} = -\big(
  \partial_{1}^2 + \partial_{2}^2 + \partial_{3}^2 \big)\,, \quad \bar
  \partial_{(3)} = 1\, .
\end{equation}
Using the above expressions, one can compute the building blocks
$\mathsf{f}_{\cal H}^{6,(k)}$ as
\begin{eqnarray}
  \mathsf{f}_{\cal H}^{6,(0)}(\beta) & = & \Big[ 1 + \tfrac{1}{135}\,
    \sinh^2\tfrac\beta2\, (111- 23\cosh\b +2\cosh 2\b)\,
    (\partial_{1}^2 + \partial_{3}^2 + \partial_{3}^2) \nn && \quad +
    \tfrac{2}{27}\, \sinh^4\tfrac\beta2\, (\cosh \b-4)\,
    (\partial_{1}^4 + \partial_{2}^4 + \partial_{3}^4 -
    6\,\partial_{1}^2\, \partial_{2}^2 -6\, \partial_{1}^2
    \,\partial_{3}^2 -6\, \partial_{2}^2 \,\partial_{3}^2) \nonumber
    \\ &&\quad + \tfrac{4}{135}\, \sinh^6\tfrac\beta2\,
    (\partial_{1}^6 + \partial_{2}^6 + \partial_{3}^6
    -15\,\partial_{1}^2\,\partial_{2}^4
    -15\,\partial_{1}^2\,\partial_{3}^4 -15\,
    \partial_{2}^2\,\partial_{1}^4 -15\,
    \partial_{2}^2\,\partial_{3}^4 \nn && \qquad \qquad \qquad \quad
    -15 \,\partial_{3}^2\,\partial_{1}^4 -15
    \,\partial_{3}^2\,\partial_{2}^4 +
    270\,\partial_{1}^2\,\partial_{2}^2\,\partial_{3}^2)\Big]\,
  \chi^{so(2,6)}_{\cal
    H}(\beta,\vec\alpha)\big|_{\vec\alpha=\vec0}\,,\qquad
\end{eqnarray}
\begin{eqnarray}
  \mathsf{f}_{\cal H}^{6,(1)}(\beta) & = & -\tfrac{1}{90}\,
  \sinh^2\tfrac\beta2\, \Big[ 111- 23\cosh\b +2\cosh 2\b
    -20\sinh^2\tfrac\beta2\,(\cosh \b-4)\, (\partial_{ 1}^2 +
    \partial_{2}^2 + \partial_{ 3}^2) \nn && -20\sinh^4\tfrac\beta2
    (\partial_{ 1}^4 + \partial_{ 2}^4 + \partial_{ 3}^4 - 6\,
    \partial_{ 1}^2\,\partial_{ 2}^2 - 6\, \partial_{ 1}^2\,\partial_{
      3}^2 - 6 \,\partial_{ 2}^2\,\partial_{ 3}^2) \Big]\,
  \chi^{so(2,6)}_{\cal
    H}(\beta,\vec\alpha)\big|_{\vec\alpha=\vec0}\,,\qquad
    \end{eqnarray}
\begin{eqnarray}
  \mathsf{f}_{\cal H}^{6,(2)}(\beta) & = & -\tfrac16\,
  \sinh^4\tfrac\beta2\, \Big[ \cosh \b-4\, - 2\, \sinh^2\tfrac\beta2\,
    (\partial^2_{ 1} + \partial^2_{ 2} + \partial^2_{ 3}) \Big]\,
  \chi^{so(2,6)}_{\cal H}(\beta,
  \vec\alpha)\big|_{\vec\alpha=\vec0}\,, \qquad
\end{eqnarray}
\begin{equation}
  \mathsf{f}_{\cal H}^{6,(3)}(\beta) = -\tfrac16\,
  \sinh^6\tfrac\beta2\, \chi^{so(2,6)}_{\cal H}(\beta, \vec0)\, .
\end{equation}
The zeta function can be readily derived using \eqref{zeta_f_even}. 

\section{Summary and Conclusion}
\label{sec: conclusion}
In this work, we derived the character integral representation of zeta
function (CIRZ) in arbitrary dimensions.  We started with a brief
review of the AdS zeta functions in
\hyperref[sec:zeta_function_AdS]{Section \ref{sec:zeta_function_AdS}},
which include its definition and some interesting identities. In
\hyperref[sec:CIRZ]{Section \ref{sec:CIRZ}}, we expressed the CIRZ
formula in terms of contour integrals in arbitrary dimensions. In \hyperref[sec:derivative_expansion]{Section
  \ref{sec:derivative_expansion}}, we derived a different CIRZ formula
in terms of derivatives, generalizing the previous derivative
expressions for AdS$_4$ and AdS$_5$ \cite{Bae:2016rgm}
to AdS$_{3}$, AdS$_6$ and AdS$_7$. This procedure also
clearly generalizes to arbitrary dimensions.

As outlined in the Introduction, the CIRZ is particularly useful to deal with theories
with an infinite spectrum. When the spectrum can be captured by some CFT data, 
as is the case for partially massless higher-spin theories \cite{Bekaert:2013zya}, 
one can compute the free energy of the theory without necessarily knowing the detailed decomposition of the spectrum.
This will be done in the companion paper \cite{part_ii}, where we will 
establish the matching of the one-loop corrections of partially
massless higher-spin gravities with the $1/N$ corrections of the dual
CFTs in any dimensions. The CIRZ method could also prove efficient in computing one-loop effects of a Kaluza-Klein tower. Indeed, the corresponding spectrum is obtained from the branching rule of the higher-dimensional field, and hence should be fully encompassed by the character of the latter.

\acknowledgments

  The research of T.B., E.J. and W.L.  was supported by the National
  Research Foundation (Korea) through the grant 2014R1A6A3A04056670.
  S.L.'s work is supported by the Simons Foundation grant 488637 
  (Simons Collaboration on the Non-perturbative bootstrap)
  and the project CERN/FIS-PAR/0019/2017.
  Centro de Fisica do Porto is partially funded by the 
  Foundation for Science and Technology of Portugal (FCT).

\appendix
\section{Character identities}
\label{app:character_stuff}
In this Appendix we shall derive the identity
\eqref{prop_character_so2+d} used in the main text. The character
formula for the $so(d+2)$ irrep labelled by the highest weight
$(s_0,\Y)$, with $\Y=(s_1,\dots,s_r)$, reads
\begin{equation}
  \chi^{so(d+2)}_{(s_0,\Y)}(\bm \alpha) = \frac{\mathsf
    N_{(s_0,\Y)}^{(d+2)}(\bm \alpha)}{\mathsf D^{(d+2)}(\bm
    \alpha)}\,,
\end{equation}
with the denominator
\begin{equation}
\mathsf D^{(d+2)}(\bm \alpha) = \prod_{0 \leqslant i < j \leqslant
      r} 2 \,( \cos\alpha_i - \cos\alpha_j)\times \left\{
  \begin{aligned}
   1 \qquad \quad & \qquad [d=2r]\\ \prod_{k=0}^r
   2\,i\,\sin\frac{\alpha_k}2\, &\qquad [d=2r+1]
  \end{aligned}
  \right.,
  \label{denominator}
\end{equation}
and the numerator
\begin{equation}\label{num_char_1}
  \mathsf N_{(s_0,\Y)}^{(d+2)}(\bm \alpha) = \left\{
  \begin{aligned}
   \frac12 \Big(\det\big[2i \sin(\alpha_i\,\ell_j)\big] +
   \det\big[2\cos(\alpha_i\,\ell_j)\big] \Big) & \qquad
           [d=2r]\\ \det\big[2i
             \sin(\alpha_i\,\ell_j)\big]\qquad\qquad\quad & \qquad
           [d=2r+1]
  \end{aligned}
  \right..
\end{equation}
Here, $\ell_j:=s_j+\frac d2-j$ and $\det[a_{ij}]$ denotes the determinant 
of a matrix whose matrix element is $a_{ij}$.
The indices $i, j$ range from $0$ to $r$.
The determinants appearing in  \eqref{num_char_1} 
may be expanded as an alternating sum of minors:
\begin{equation}\label{num_char_2}
\mathsf N_{(s_0,\Y)}^{(d+2)}(\bm \alpha) = \sum_{k=0}^r (-1)^k \times
\left\lbrace
  \begin{aligned}
    &\quad \begin{aligned} &\Big( i\sin(\alpha_k\,\ell_0)\, {\det}_k
      \big[2i\sin(\alpha_i\,\ell_j)\big] \\ &\quad +
      \cos(\alpha_k\,\ell_0)\, {\det}_k
      \big[2\cos(\alpha_i\,\ell_j)\big] \Big)
    \end{aligned}\, & \qquad   [d=2r]\\ 
   &\quad 2\,i\, \sin(\alpha_k\,\ell_0)\, {\det}_k\big[2i
      \sin(\alpha_i\,\ell_j)\big] & \qquad [d=2r+1]
  \end{aligned} 
  \right..
\end{equation}
The above expression can be recast into recursive formulae in $d$ as 
\begin{equation}\label{num_char_3}
  \mathsf N_{(s_0,\Y)}^{(d+2)}(\bm \alpha) =\sum_{k=0}^r\,
  (-1)^k\times\left\{
  \begin{aligned}
    \Big( e^{i\alpha_k\,\ell_0}\, \mathsf N^{(d)}_{\Y_+}(\vec
    \alpha_k) + e^{-i\alpha_k\,\ell_0}\, \mathsf N^{(d)}_{\Y_-}(\vec
    \alpha_k) \Big)\,&\qquad \qquad [d=2r]\\ 2i\,
    \sin(\alpha_k\,\ell_0)\, \mathsf N^{(d)}_{\Y}(\vec \alpha_k)
    \qquad \quad\, & \qquad [d=2r+1]
  \end{aligned}
  \right..
\end{equation}
For $d=2r+1$, \eqref{num_char_3} is straightforward to prove. 
For $d=2r$, the formula \eqref{num_char_2} can be first expanded as
\begin{eqnarray}
  \mathsf N_{(s_0,\Y)}^{(d+2)}(\bm \alpha) & = & \frac12\,\sum_{k=0}^r (-1)^k
  \Big( e^{i\alpha_k\,\ell_0}\, \big( {\det}_k
    [2i\sin(\alpha_i\,\ell_j)] + {\det}_k [2\cos(\alpha_i\,\ell_j)]
    \big) \nn && \qquad \qquad\quad +\, e^{-i\alpha_k\,\ell_0}\, \big(
    {\det}_k [2\cos(\alpha_i\,\ell_j)] - {\det}_k
    [2i\sin(\alpha_i\,\ell_j)] \big) \Big)\, ,
\end{eqnarray}
then the relative minus sign in the second line can be absorbed into
the determinant by changing the last column from $\sin(\alpha_i\,
\ell_r)$ to $\sin(-\alpha_i\,\ell_r)$. \\

Similarly, one can decompose the denominator of an $so(d+2)$ character
as a multiple of the denominator of an $so(d)$ character expressed in
terms of $\vec \alpha_k$:
\begin{equation}\label{D to D}
  \mathsf D^{(d+2)}(\bm \alpha) = (-1)^k\, \mathsf D^{(d)}(\vec
  \alpha_k)\, \prod_{\substack{0 \leqslant i \leqslant r\\ i \neq k}}
  2 \,(\cos\alpha_k-\cos\alpha_i) \times \left\{
  \begin{aligned}
    1 \quad & \qquad [d=2r]\\ 2i \sin\tfrac{\alpha_k}{2} & \quad
    [d=2r+1]
  \end{aligned}
  \right.,
\end{equation}
where the overall factor $(-1)^k$ comes from re-ordering part of the
product containing the $\alpha_k$ variables in $\mathsf D^{(d+2)}(\bm
\alpha)$. Note that \eqref{D to D} can be expressed as
\begin{equation}\label{DD}
  \mathsf D^{(d+2)}(\bm \alpha) = (-1)^k \, \mathsf D^{(d)}(\vec
  \alpha_k)\, \frac{(-1)^d\, e^{-i\alpha_k\,d/2}}{\Pd d (\alpha_k;\vec
    \alpha_k)}\, .
\end{equation}
by using the definition of the function $\Pd d$ defined in
\eqref{def_Pd}. Combining \eqref{num_char_3} and \eqref{DD}, we can
finally express the $so(d+2)$ characters in terms of $so(d)$
characters as
\begin{eqnarray}
  \chi^{so(d+2)}_{(s_0,\Y)}(\bm \alpha) = \sum_{k=0}^r\, \Pd
    d (\alpha_k;\vec \alpha_k) \left\{ 
  \begin{aligned}
   \left( e^{-i\alpha_k\,s_0}\, \chi^{so(d)}_{\Y_-}(\vec \alpha_k) +
   e^{i\alpha_k(s_0+d)}\, \chi^{so(d)}_{\Y_+}(\vec \alpha_k)\right) &
   \quad [d=2r] \\ \big( e^{-i\alpha_k\,s_0} -
   e^{i\alpha_k(s_0+d)}\big) \, \chi^{so(d)}_\Y(\vec \alpha_k)\,
   \qquad & \quad [d=2r+1]
  \end{aligned}
  \right..\nn
\end{eqnarray}

\section{Generalized L'H\^opital's rule}
\label{sec: app B}
In this Appendix, we will discuss some technical details of the
generalized L'H\^opital's rule \eqref{eq:charlimit} (see
e.g. \cite{Hall2003} for a pedagogical introduction) which are crucial
in obtaining the derivative expression of CIRZ in
\hyperref[sec:derivative_expansion]{Section
  \ref{sec:derivative_expansion}}. The differential operator
$\mathscr{D}_{\Phi^{d+2}}$ is defined as
\begin{equation}
  \mathscr{D}_{\Phi^{d+2}} = \prod_{\theta \in \Phi^{d+2}}
  \partial_\theta\, .
  \label{diff_op_so2+d}
\end{equation}
In the orthonormal basis, the set of positive roots of $so(d+2)$ reads
\begin{equation}
  \Phi^{d+2} = \left\{
  \begin{aligned}
    \{ \mathsf e_i \pm \mathsf e_j\, , 0\leqslant i < j \leqslant r
    \}\, \qquad \qquad & \qquad [d=2r]\\ \{ \mathsf e_i \pm \mathsf
    e_j\, , 0\leqslant i < j \leqslant r \} \cup \{ \mathsf e_k\, , 0
    \leqslant k \leqslant r \}\, & \qquad [d=2r+1]
  \end{aligned}
  \right.\,,
\end{equation}
where $\{\mathsf e_k\}_{k=0,\,\dots,\,r}$ is a basis of unit
orthonormal vector of $\mathbb R^{r+1}$. As a consequence, when acting
on an $so(d+2)$ character, we have
\begin{equation}
  \partial_\theta = \tfrac12 (\partial_{\alpha_i} \pm
  \partial_{\alpha_j}) \quad \text{for} \quad \theta = \mathsf e_i \pm
  \mathsf e_j\,, \quad \text{and} \quad \partial_\theta =
  \partial_{\alpha_k} \quad \text{for} \quad \theta=\mathsf e_k\,.
\end{equation}
From now on, we will use the notation
$\partial_i=\partial_{\alpha_i}$.  Since the differential operator
\eqref{diff_op_so2+d} acts on both the numerator and denominator, we
can rescale this operator to eliminate the factor $\tfrac 1 2$ from
$\partial_\theta$ with $\theta = \mathsf e_i \pm \mathsf e_j$.  The
differential operator $\mathscr D_{\Phi^{d+2}}$ then takes the form
\eqref{D def}.

\subsection{Computing the denominator}
\label{sec: den}
According to \eqref{denominator}, the denominator of the $so(d+2)$
character depends on the dimension $d+2$, but not the highest weight
$(s_0,\Y)$ of the representation. As a result, the denominator of the
generalized L'H\^opital's rule is simply a $d$-dependent constant,
which is denoted $c_d$ in \eqref{c d}. We will review a part of the
derivation of the Weyl dimension formula, which can be found in
e.g. \cite{Hall2003} and enables us to obtain an explicit formula for
$c_d$. First, recall that the denominator of an $so(d+2)$ character
can be expressed as the product,
\begin{equation}
  \mathsf D^{(d+2)} = \prod_{\theta \in \Phi^{d+2}}
  (e^{\theta/2}-e^{-\theta/2})\,.
\end{equation}
One can then show that the action of the differential operator
\eqref{diff_op_so2+d} on $\mathsf D^{(d+2)}$ reads
\begin{equation}
  \mathscr D_{\Phi^{d+2}} \prod_{\theta \in \Phi^{d+2}} (e^{\theta/2}-e^{-\theta/2}) =
  |\mathcal{W}_{d+2}| \prod_{\theta\in\Phi^{d+2}} \langle \theta,
  \rho \rangle\, ,
\end{equation}
where $\r$ is the $so(d+2)$ Weyl vector:
\begin{equation}
  \rho:=\tfrac12 \sum_{\theta\in\Phi^{d+2}}\, \theta,
\end{equation}
and $|\mathcal{W}_{d+2}|$ is the cardinal of
the Weyl group of $so(d+2)$
\begin{equation}
  |\mathcal{W}_{d+2}| = \left\{ 
  \begin{aligned}
    2^r\, (r+1)!\,\,\, & \qquad \qquad [d=2r]\\ 2^{r+1}\, (r+1)!\, &
    \qquad \qquad [d=2r+1]
  \end{aligned}
  \right.\,,
\end{equation}
and $\langle\, ,\rangle$ denotes the inner product on the 
root space. 
Since in this paper we have used  $e^{i\theta}$,
instead of $e^\theta$, for the variables of the $so(d+2)$ character,
the constant $c_d$ is given by
\begin{equation}
  c_d = i^{|\Phi^{d+2}|} |\mathcal{W}_{d+2}|
  \prod_{\theta\in\Phi^{d+2}} \langle \theta, \rho \rangle,
\end{equation}
with $|\Phi^{d+2}|$ being the number of positive roots of $so(d+2)$, 
\begin{equation}
  |\Phi^{d+2}| = \left\{
  \begin{aligned}
    r(r+1)\, & \qquad [d=2r]\\ (r+1)^2\, & \qquad [d=2r+1]
  \end{aligned}
  \right.\,.
\end{equation}
Since the Weyl vector of $so(d+2)$ in the orthonormal
basis is
\begin{equation}
  \rho = \sum_{k=0}^r (\tfrac d2 - k)\, \mathsf e_k,
\end{equation}
we finally obtain $c_d$ as given in \eqref{formula_c}.

\subsection{Simplifying the numerator}
\label{sec: num}
Using \eqref{num_char_3}, we can express $\mathsf N^{(d+2)}$ in terms
of $\mathsf N^{(d)}$, then the numerator of the generalized
L'H\^opital's rule becomes
\begin{equation}
  \mathscr{D}_{\Phi^{d+2}}\, \mathsf{N}_{(s_0,\Y)}^{(d+2)}(\bm \alpha)
  \big|_{\bm \alpha = \bm 0}= 2\, \sum_{k=0}^r \,(-1)^{k+d}\,
  \mathscr{D}_{\Phi^{d+2}} \Big[ \sigma_d(\alpha_k u)\, \mathsf
    N_\Y^{(d)}(\vec \alpha_k) \Big]\Big|_{\bm \alpha=\bm 0}
\end{equation}
with
\begin{equation}
  \sigma_d(\alpha_k\, u) = \left\{
  \begin{aligned}
    \cosh(\alpha_k\,u)\, & \qquad [d=2r] \\ \sinh(\alpha_k\, u)\, &
    \qquad [d=2r+1]
  \end{aligned}
  \right.\,.
  \label{sigma}
\end{equation}
Here, we took advantage of the facts that
\begin{equation}
  \mathsf N_{\Y_-}^{(d)}(\alpha_1, \dots, \alpha_{k-1}, -\alpha_k,
  \alpha_{k+1}, \dots, \alpha_r) = \mathsf N_{\Y_+}^{(d)}(\alpha_1,
  \dots, \alpha_{k-1}, \alpha_k, \alpha_{k+1}, \dots, \alpha_r)\, ,
\end{equation}
and that for $d=2r$ the differential operator
$\mathscr{D}_{\Phi^{d+2}}$ is invariant under the $\alpha_k
\rightarrow -\alpha_k$ transformation for any $k=0, \dots, r$.  The
differential operator $\mathscr{D}_{\Phi^{d+2}}$ can then be rewritten
as:
\begin{equation}
  \mathscr{D}_{\Phi^{d+2}} = (-1)^k\, \mathscr{D}_{\Phi^d \rvert
    k}\, \prod_{\substack{0 \leqslant i \leqslant r \\ i \neq k}}
  (\partial_k^2 - \partial_i^2)\, \times \, \left\{
  \begin{aligned}
    1\,\, & \qquad [d=2r]\\ \partial_k\, & \qquad [d=2r+1]
  \end{aligned}
  \right.\,,
  \label{diff_op_sod+2}
\end{equation}
where $\mathscr{D}_{\Phi^d \rvert k}$ is expressed in
terms of the unit vectors $\{ \mathsf e_0, \dots, \mathsf e_{k-1},
\mathsf e_{k+1}, \dots, \mathsf e_r \}$ of $\mathbb R^r$. Expanding
the last product, we can then isolate the derivative with respect to $\alpha_k$:
\begin{equation}
  \prod_{\substack{0 \leqslant i \leqslant r \\ i \neq k}}
  (\partial_k^2 - \partial_i^2) = \sum_{n=0}^r \partial_k^{2n} \,\bar
  \partial_{(n|k)}\, ,
\end{equation}
with
\begin{equation}
  \bar \partial_{(n|k)} = (-1)^{r-n}\, \sum_{\substack{0\le i_1<i_2<\dots<
    i_{r-n}\leq r\\ i_j \neq k}}
  \partial_{i_1}^2 \dots \partial_{i_{r-n}}^2\,.
\end{equation}
In particular, we
have
\begin{equation}
  \bar \partial_{(r|k)} = 1\, , \quad \bar \partial_{(r-1|k)} = -
  \sum_{\substack{0 \leqslant i \leqslant r \\ i \neq k}}
  \partial_i^2\, , \quad \text{and} \quad \bar \partial_{(0|k)} =
  (-1)^r \prod_{\substack{0 \leqslant i \leqslant r \\ i \neq k}}
  \partial_i^2\, .
\end{equation}
With this decomposition at hand, we can write
\be
  \mathscr{D}_{\Phi^{d+2}} \Big[  \sigma_d(\alpha_k u) \, 
    \mathsf  N_\Y^{(d)}(\vec \alpha_k) \Big]\Big|_{\bm
    \alpha=\bm 0} \nn
     =  (-1)^k\, u^{d-2r}\, \sum_{n=0}^r u^{2n}\,
  \bar \partial_{(n|k)}\, \mathscr{D}_{\Phi^d \rvert k}\, 
   \mathsf  N_\Y^{(d)}(\vec \alpha_k) \Big|_{\bm
    \alpha=\bm 0}\, ,
\ee
where we have used the fact that the dependency on the angle
$\alpha_k$ is confined to the function $\sigma_d(\alpha_k u)$. 
On top of that, this enable us to extract the $u$ dependent part 
according to $\partial_k$. 
Notice finally that the remaining summands involve 
all variables $\{\alpha_i\}_{i=0,\dots,r}$ except for $\alpha_k$. 
As a consequence, the terms
\begin{equation}
  \bar \partial_{(n|k)} \mathscr{D}_{\Phi^d \rvert k} 
   \mathsf  N_\Y^{(d)}(\vec \alpha_k) \Big|_{\bm
    \alpha=\bm 0}\,
\end{equation}
all produce the same contribution for different $k=0,\dots,r$. We can
therefore drop the subscript $k$ in the above expressions.  In the
end, we obtain \eqref{DN} with \eqref{differential_operator}, where
the factor $u^{d-2r}$ is from the additional $\partial_k$ in
\eqref{diff_op_sod+2} for $d=2r+1$.

\bibliographystyle{JHEP}
\bibliography{biblio}

\end{document}